# A CAD-BASED TOOL FOR FAULT TOLERANT DISTRIBUTED EMBEDDED SYSTEMS

Mahmoud I. Banat[1], Belal H. Sababha[2]*, Senior Member, IEEE, and Sami Al-Hamdan[1]
[1]Computer Engineering Department, Yarmouk University, Irbid, Jordan
[2]Computer Engineering Department, Princess Sumaya University for Technology, Amman 11941, Jordan
*Corresponding author: Belal H. Sababha (e-mail: b.sababha@psut.edu.jo).

ABSTRACT

Reliability and availability analysis are essential in dependable critical embedded systems. The classical implementation of dependability for an embedded system relies on merging both fundamental structures with the required dependability techniques to form one composite structure. The separation of the basic system components from the dependability components, reduces complexity and improves the design. The goal of this work is to assist implementing reconfiguration-based fault tolerance in safety-critical embedded systems applications. The primary intention is to reduce the repair time in order to enhance fault tolerance and produce dependable embedded systems. The proposed solution is a dedicated CAD-tool designed to generate a reference strategy for the system manager of a distributed embedded system to control and automatically reconfigure the processing elements of the system. The proposed tool auto-generates program codes to be executed by a system manager to govern the DES. It also computes different reliability solutions with necessary supporting calculated parameters and graphs sorted to support the fault tolerance design of the system. The proposed tool can be used to simulate possible configurations based on the desired degrees of faults and system reliability. The graphical interface of the tool is unique and hides the complexity of the systems underneath. A comparison with a similar tool is presented.

## 1. INTRODUCTION

Distributed Embedded Systems (DES) are increasingly becoming safety-critical and permeate lives. These applications are very diverse covering aircraft, cars, finance machines, electronic appliances, networking devices, and military equipment. Reliance on Embedded Systems (ES) have become a main part of today's systems and devices. The increasing number of interconnected distributed devices is increasing the complexity of ensuring and maintaining dependability. Safety-critical applications with embedded computing systems are all over. Continuous dependability evaluation for these applications is vital to achieve the required satisfaction degree of stability, availability, security and integrity.

Fault tolerance techniques aim to enhance the dependability of a system, and they are usually combined with other fault detection and isolation methods such as removal, prevention, and forecasting. Reconfiguration-Based Fault tolerance or graceful degradation as sometimes called, enhance manufacturing yield and resilience towards malfunctions. Fault tolerance is vital in ES design. ESs are intended to be implemented in harsh environments with very strict resources and conditions, such as hazardous circumstances, underwater spaces, lack of power conditions, and time constrains.

Implementing auto-code generation during the design process could enhance embedded systems dependability by avoiding design errors and minimizing their origins and causes. Auto-code generation also increases the average time between failures during operation and increases the total dependability of the system by reducing the required maintenance time in case of failures.

This paper introduces a CAD tool designed to automatically generate source codes from dependency graph specifications. Traditional hard coding methods are stationary and inflexible, because the designer has to be aware of all failure possibilities and be conscious to a code failure mitigation from the beginning, such that if any change is done, then a total modification may be needed. Since fault tolerance code is mixed with the design of the system, professional embedded system designers are expected to act properly when they face various cases of difficulty. Unfortunately, a few number of competent designers make the traditional coding process poor and slow [1]. Therefore, there is demand to reduce drawbacks that are caused by hardcoding fault tolerance techniques used in the ES filed. If a separate system is used to analyze the different design options, classify them, and select the best reactions for a condition, this will help in evaluating different assignments without getting involved in the high cost inquiries when using one composite design.

In this work, Computer Aided Design (CAD) is used to generate, modify, analyze and improve the design of FT distributed embedded systems, and particularly for fault tolerant embedded systems.

Recent dependent DES frameworks suffer from many shortcomings and problems. For example, there is a lack of automated dynamic reconfiguration approaches to substitute traditional labor-intensive pre-coding fault tolerance techniques. Human mistakes affect negatively the dependency of the DES. Some mistakes might be disastrous specifically when they are related to safety critical applications. The use of manual FT design and coding effects the production time and cost significantly, especially when perceptible efforts are paid on the designs and their variations and adjustments analysis, which take place all the way during the manufacturing process.

Manual FT DES reconfiguration methods are not scalable and require frequent interaction from the designer and the operator during the design stage and during runtime. Obviously, manual reconfiguration can only support limited quantities of Processing Elements (PE) of an Embedded System.

This work contributes to the automation of fault tolerant embedded systems through presenting a new CAD-based tool. The here presented tool is based on the ARDEA framework (Automatically Reconfiguring Distributed Embedded control Architecture) [2].

The proposed CAD tool is intended for reconfiguration-based FT Design in DES. It can automatically communicate with the FT system manager to implement the required reconfiguration options to be taken when faults occur.

The rest of the paper is organized as follows. Section 2 gives the required background about fault tolerance. Section 3 reviews the existing research and literature related to FT DES frameworks. Section 4 introduces the proposed fault tolerance design CAD-based tool. Section 5, presents the results and outcomes along with a comparison with other dependency tools. Finally, section 6 concludes the paper and presents possible directions for future research.

## 2. BACKGROUND

Dependability is “the ability of a system to deliver its intended level of service to its users” [3]. Dependability expresses the system's availability, reliability and its maintainability [4]. Which means dependability is a measure of a system’s resilience in front of faults and their implications before totally failing. Fundamental characteristics of dependability are attributes, impairment, and means [5]. Attributes describe the properties required from a system to be dependable. Impairments are the threats that negatively affect the system dependability and form risks for a system dependability. Means are the techniques that are used to affect positively the development of dependable systems and increase systems reliance.

Availability expresses the fraction of time during which a system meets its required specification. Availability is important when continuous service is significant, but it can be delayed or rejected [6]. It is defined by [7] as “the readiness for correct service”. Eq. 1 expresses the mean value of the instantaneous availability function over the period (0, T], [8]:

$$A(T) = \frac{1}{T}\int_0^T A(t).\mathrm{d}t \qquad \text{Eq. 1}$$

Eq. 2 shows how reliability and Mean Time to Failure (MTTF) can be derived from the basic notion of failure rate. MTTF in a sample of identical initially perfect systems is defined as the expected value of Life $t_f$

$$\mathrm{MTTF} = \mathrm{E}(t_f) = \int_0^\infty \mathrm{R}(t)\mathrm{d}t = \frac{1}{\lambda} \qquad \text{Eq. 2}$$

Some typical system availabilities and their ranking can be found at [9].

Another required attribute of a dependable system is reliability. "Reliability R(t) of a system at time t is a measure of correct service continuous delivery" [7], or survival probability in interval [0, t], assuming that the system was operating appropriately at time 0.

Reliability reflects the probability of success. On the other hand, unreliability Q(t) of a system at time t, can be defined as the probability of a system failure in the interval [0, t], given that it was acting correctly at time 0 [5]. Eq. 3 [5] shows the complementary relation between reliability and unreliability (Q), and Eq. 4 [5] illustrates the probabilistic expression for the Reliability

$$R_F(t) + Q_F(t) = 1 \qquad \text{Eq. 3}$$

$$R_F(t) = P(t_{init} \leq t < t_f, \forall f \in F) \qquad \text{Eq. 4}$$

P is the reliability probability, $t_{init}$ is the time of introduction the system to service, $t_f$ is time of the first failure f drawn from all failures domain F.

A large number of models have been proposed to evaluate and describe the dependability attributes [10, 11, 12, 13]. The dependency evaluation models are also referred to as probabilistic models as they abstract availability probabilistic properties of the system. There are many probabilistic and statistical models used in reliability [14]. Exponential distribution is chosen in this work because it plays an essential role in modeling reliability as it is the only continuous distribution with a constant hazard function [14]. Eq. 5 expresses the component reliability R(t), where λ is the failure rate of the component under study and forms the hazard function for the distribution function.

$$R(t) = e^{-\lambda t} \qquad \text{Eq. 5}$$

For a system with multiple components, if any component failure causes the failure of the entire system, then the system is then called a series system. In a serial system, and assuming that the failure rate (λ) of every single component K is statistically independent, then the reliability of the component RK(t) with respect to time is expressed by Eq. 6 [5].

$$R_K(t) = e^{-\lambda_K t} \qquad \text{Eq. 6}$$

The overall system reliability in case of complex system are subject to the discrete units that build that system. Eq. 7 and Eq. 8 [5, 8] express the reliability of a serially connected components:

$$R_{ser}(t) = e^{-t(\sum_{k=1}^{n} \lambda_k)} \qquad \text{Eq. 7}$$

$$R_{ser}(t) = R_1(t) \times R_2(t) \times R_3(t) \times \ldots \times R_n(t) \qquad \text{Eq. 8}$$

Eq. 9 [8] illustrates that the serial failure rate will be the cumulative failure rate $\lambda_{ser}$ of every component along the serial path, therefore a reliable serial system is very difficult to build and expensive [9].

$$\lambda_{ser} = \sum_{k=1}^{n} \lambda_k \qquad \text{Eq. 9}$$

In parallel systems, the system fails only if all components of the system fail. Parallel Connected Components' unreliability Qk(t) is expressed by Eq. 10 from [8]:

$$Q_K(t) = 1 - e^{-\lambda_k t} \quad \text{Eq. 10}$$

When failure rates of components are statistically independent, Eq. 11 expresses the unreliability $Q_{par}(t)$. However, it is out of scope of this research to discuss the statistically dependent components.

$$Q_{par}(t) = \prod_{k=1}^{n} Q_k(t) \quad \text{Eq. 11}$$

The total system Unreliability will be ruled by Eq. 12 [5] [8]:

$$R_{par}(t) = 1 - \prod_{i=k}^{n} \left(1 - R_k(t)\right) \quad \text{Eq. 12}$$

Design errors could be the first source of faults, because of problems in manufacturing, installation, configuration, and wiring. Alternatively, due to environmental external disturbances or harsh conditions. System misuse is one of fault sources, or mechanical causes of deterioration like the cases of wear, fatigue, or corrosion, or like shock, overload, etc. [16]. Electronic hardware problems could occur because of bad electronic hardware fabrication. Design defects, operating environment, manufacturing defects are factors that cause malfunctions of a system. Some error sources, their phases, and methods of detecting them can be found at [16]. Sample failure rates for automotive embedded systems expressed in failures per million operating hours can be found at [9].

## 3. LITERATURE REVIEW

Fault Tolerant Distributed Embedded Systems (FT DES) are most often studied in terms of system reliability and availability. In the literature, there are numerous examples of new models to perform this type of FT. Many different approaches have been proposed to tackle down the complication of the FT DES. Different approaches entail setting new parameters. Relevant work is discussed in this section.

Since 1979, there have been many attempts to establish a framework for DES dependability. COFTA [17] was presented as a model to evaluate reliability and availability of the architecture during co-synthesis in both software and hardware [17]. COFTA allowed FT based task clustering, a small number of extra processing elements added to get the best error recovery topology, and assertions shared to reduce the FT overhead.

[18] discussed the concept of a specialized N-modular redundancy (SNMR) framework in the design. The Amaranth Project at Carnegie Mellon searched and developed a framework for allocating shared resources trying to offer probabilistic assurances of service and support quality of service (QoS) [6]. Egida, is another extensible toolkit designed to facilitate rapid implementation of arbitrary rollback recovery protocols with minimal programming effort proposed by Rao [19].

Some approaches look at the FT from security perspective. Kim [20], examined how to increase FT by preventing unauthorized users either from accessing protected information or from denial-of-service attacks.

Shelton divided the structure of the system architecture into subsystems to avoid exponential complexity of component combinations [21]. The framework designed to view each subsystem as a configuration of components, when components fail or are repaired, the subsystems will change.

For high reliability, [22] believed that hybrid FT model must be tuned to the statistics of faults expected, types and rates. By determine the dominant cause of failure; one can address the appropriate algorithm at design time. Which yields to a significant reliability gain. In his design, he examined the membership and clock synchronization reliability [22].

Some methods encompass machine-learning aspects, [23] used forward and backward neural networks in recovery mechanism. The authors built their technique on execution of design with variants settings and different hardware, then assigning weights to the results produced.

These studies hinted that FT might be implemented based on component-based attribution style. A later experiment showed that perfectionism level could be correlated with interdependencies of resources. [24] in his dynamically reconfiguration-based architecture, the author defined modules' interdependencies using dependency graphs. If a failure occurs, a system manager tracks the status of the system and feeds software modules the required alternate implementation with different system resources. Rawashdeh's architecture supports redundant calculations, voting, and fail-over programming. ARDEA forms the graphical specifications of the Rawashdeh's framework [2].

The first time a state machine operating approach in conjunction with modal function blocks was used by Angelov *et al.* [25]. The authors used a hybrid executable model-a clocked event-driven in distributed embedded applications, and they divided the software into two blocks, actors that form the supervisory state machines, and another that form the executable components, those two blocks form a clocked event-driven, hybrid executable model.

A modular architecture approach and a system level virtualization exploitation has been conducted by Vallee to implement a proactive migration and pause/un-pause FT policies. The work compared the results by simulation [26].

Hamid, *et al.* [27], focused in their approach on considering a model as a first structural citizen, not component-based. They used Unified Modelling Language (UML), extended with some of its dedicated profiles, to provide FT services in underlying execution middleware. Replication management mechanisms enabled, and some dedicated system components added to detect faults and some added for management purposes.

One frameworks called CREAM, [28] tried to meet together lightweight memory use, fast communication inter-component, distributed processing support, transparent communication, and task execution integrity. Meanwhile, Emberson [29] addressed a search algorithm in order to optimize fault-tolerance properties; they discussed the ability to search local neighborhood and balance constraints with objectives of the design.

After that, Glass [30] analyzed degradation-aware reliability to allocate resources and function distribution. He applied data structures to utilize an efficient degradation algorithm. His methodology optimized structural and behavioral optimization of the system. However, Vadlamudi *et al.* [31] characterized the components of the system by storing lookup tables (LUTs) that store the relations between inputs and outputs, and they authorized simulation to filter false positives from their framework.

More recent research as [32] tried to address research challenges in safety evaluation of automotive electronics using virtual prototypes. Waszecki, *et al.* [33] presented an approach for the diagnosis of permanent faults in automotive Electrical and Electronic architectures. Zheng, *et al.* [34] addressed the safety of real-time embedded systems with respect to both functional and timing correctness.

Laufenberg, *et al.* [35] presented an error injection and simulation-based approach for safety analysis, the work used graph-based guidance with error effect simulation. Xu, *et al.* [36] proposed a fault-tolerant mechanism for graph and machine learning, which runs on distributed systems. They tried to shorten failure recovery times by graph processing, rather than writing checkpoints that block operators. Their mechanism focuses on unblocking checkpoints by benefiting from local logging of a distributed system and using machine learning to recover the system in a case of failures.

The authors were not able to find any work that could be utilized for fault tolerant distributed embedded systems during design time and is visual CAD-based especially with implementing the most recent advances and techniques and frameworks for FT DES. Techniques so far depend on manual reconfiguration, which works fine for a few processors, but does not scale to a large number of processors. The authors believe that the developed visual CAD-based tool will contribute to filling this gap and facilitate the design of FT DES.

## 4. PROPOSED FAULT TOLERANCE DESIGN SYSTEM

The proposed CAD-Based Fault Tolerance design system is presented in this section. The tool is to be used to design FT DESs. It provides the Fault Tolerance designer with a graphical interface to represent FT techniques, set reliabilities, analyze and visualize FT options. The proposed system also enables the export of the results, i.e. Fault tolerance computing techniques to the DES.

This tool shall facilitate the work in the FT DES field. It shall enable FT DES designers to focus on high level design requirements rather than tiny detailed requirements. The expected tool shall be general purpose, practical, user-friendly, minimal requirement of resources, can import/export configurations in compatible formats, and be scalable with high performance.

If the tool can represent the FT solutions in a manner that enables the designer to view different FT options before choosing the optimal set of FT solutions, this would be another advantage of the proposed tool. Given the ability to freely try different design options, would contribute to the optimality and perfection of the design. Moreover, other basic capabilities are important to be included in the tool, such as the capabilities of extracting out, and sorting the dependencies that guarantee the DES proper operation.

ARDEA framework [2], is considered as an attractive FT framework to be a base to design and develop the proposed tool.

### *A. ARDEA Framework*

The proposed CAD-based tool is based on ARDEA framework [2, 24]. Automatically Reconfiguring Distributed Embedded Architecture (ARDEA) is a framework for developing dependable ES based on graphical software specification [24]. ARDEA approach supports traditional fault-tolerance techniques as N-version redundant calculation, fail-over programming, voting, as well as dynamically manipulating standby redundancies. These features make ARDEA considered as an attractive for designing FT DES control applications [24]. The framework is developed around a design approach based on a graphical software specification technique [24]. Dependency graphs (DGs) are used to specify the interaction and interdependencies among modules. Individual software modules can be specified with alternate implementations that may require different system resources. The top-level architecture of ARDEA is shown in Figure 1. ARDEA is a product family architecture. Product family architectures simplify the implementation of products of the same family with varying features [37]. ARDEA assumes a parallel data pipeline architectural style for application software. A Parallel data pipeline architecture is defined in [38].

A pipeline is a data flow from sources to their destinations. ARDEA assumes no violations of exceeding the available hardware resources such as bandwidth and processor time, to avoid missing deadlines and overloading the system. Aperiodic and sporadic events are handled as periodic tasks with large periods. This outcomes pipelines to have periodic architectures making them more suitable for safety-critical and real-time applications [2, 39].

ARDEA defined dependency gates that specify the dependency of modules and data variables. Figure 2 from [2] shows the graphical representation of the four types of dependency gates. The first gate in Figure 2(a) is the OR gate; it has multi-input (n) and one output. K specifies how many inputs are desired. The AND gate shown in Figure 2(b) requires all inputs. The XOR gate shown in Figure 2(c), requires exactly one input. XOR will use this only input, but OR will use at least one input branch. The last DEMUX in Figure 2(d) has one input but more than one output to allow redundancy.

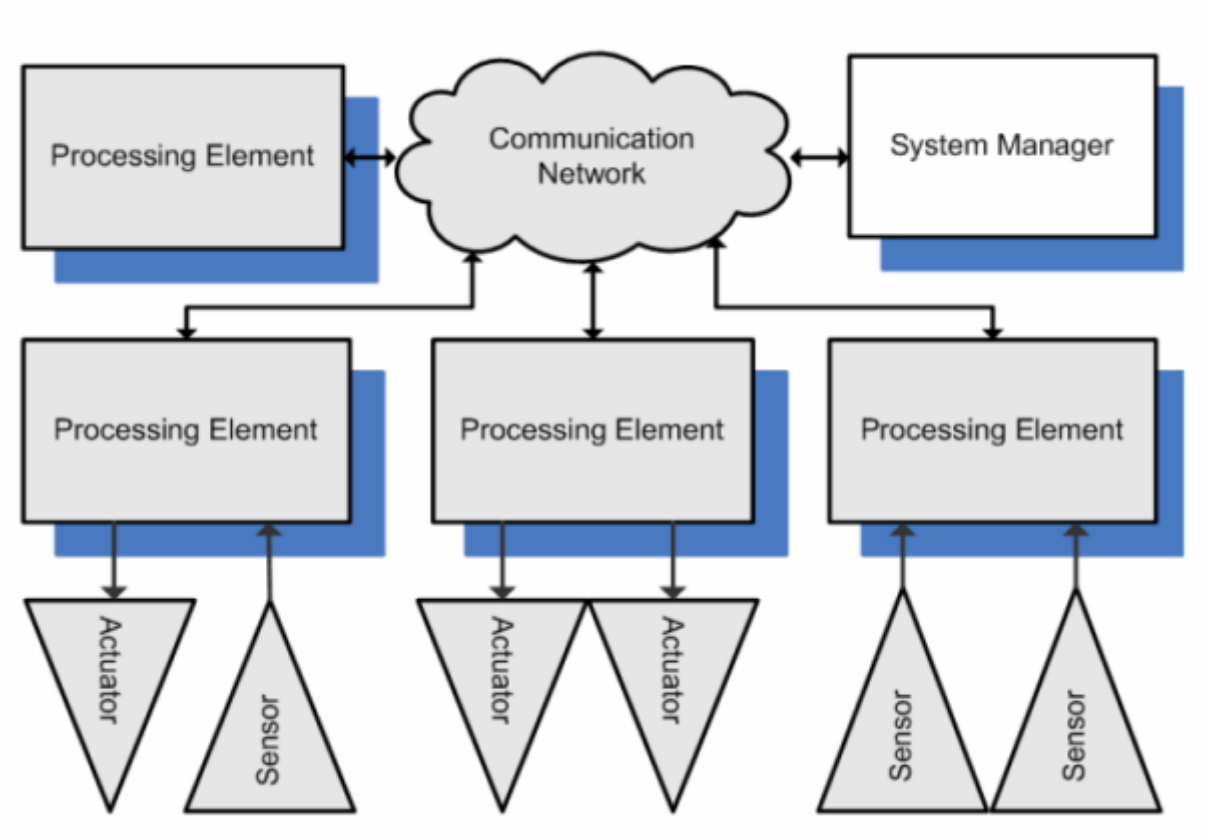

Figure 1. Hardware Block Diagram from [2].

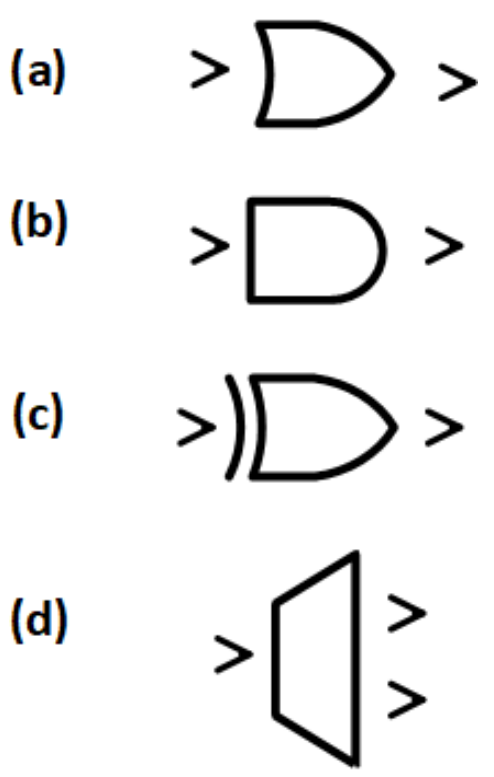

Figure 2. Dependency Gates from [2].

(a) Dependency OR. (b) Dependency AND.

(c) Dependency XOR. (d) Dependency DEMUX.

### B. *Design of the Tool*

A very high-level representation of the required tool is illustrated in Figure 3, which shows that the fault tolerant designer graphically represents the FT plan according to his experience and what the system requires, explaining the steps of the work, dependency and flow. The designer sets the policy he wants to run the system based on and sets the variables for each element of the design. Variables could be names, identifications, fault rates, Reliability, and other settings. The tool then computes the solutions and finds proper suggestions to guarantee the functionality of the system. Solutions will be exported in sort of a data file and embedded into the DES system manager mentioned in ARDEA. Data file transfer is achieved by using any file transfer protocol or using serial communication.

#### 1) BUILDING THE CAD-TOOL

The proposed system requires only a PC with any web browser. No special characteristics should the location of implementation have. The tool was developed using JavaScript language. JavaScript is a high-level, dynamic, and interpreted programming language [40]. JavaScript is one of the three-core technologies of the WWW content production, HTML and CSS form the rest. Majority of websites employ it, and all modern Web browsers support it without the need for plug-ins [40]. JavaScript is imperative and has a functional programming styles [40]. It has an API for working with contents, arrays, and regular expressions, but does not include any I/O, such as networking, storage, or graphics facilities, relying for these on the host environment in which it is embedded [40].

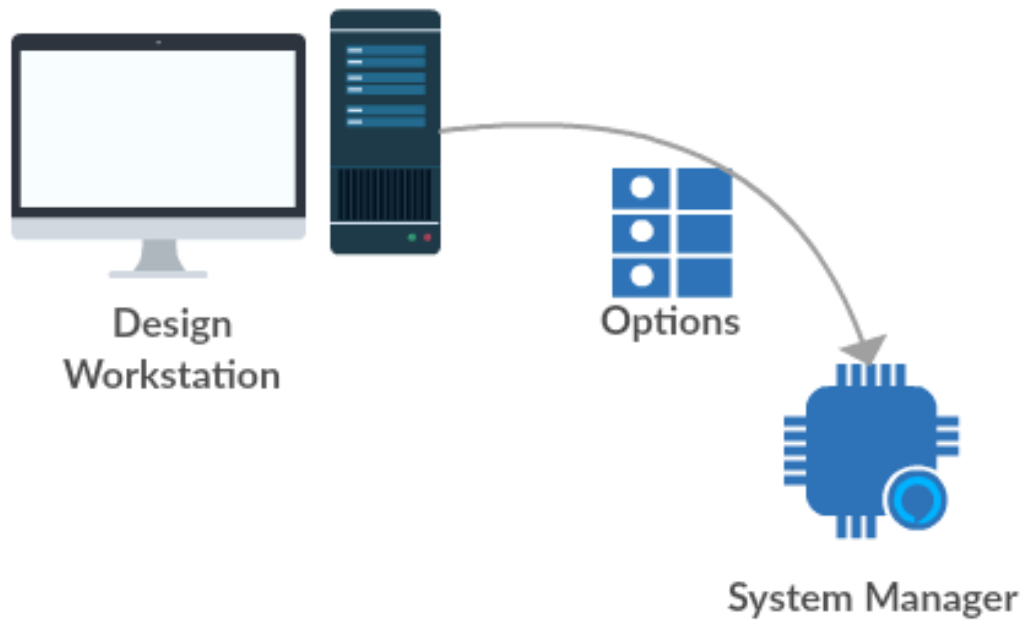

Figure 3. High-level representation of CAD-tool and embedded System manager relationship.

JavaScript is also used in environments that are not Web-based, such as PDF documents, site-specific browsers, and desktop widgets. JavaScript is also used in game development, in mobile applications, and in server-side network programming with run-time environments such as Node.js [41].

Because JavaScript code can run locally in a user's browser, rather than on a remote server, the browser can respond to user actions quickly, making an application more responsive. Furthermore, JavaScript code can detect user actions that HTML alone cannot, such as individual keystrokes.

The CAD-tool designed in this work uses a library called GoJS. GoJS is a JavaScript library for implementing custom interactive diagrams and complex visualizations across modern web browsers and platforms.

The convenience for GoJS that this library does not need extra plugins and runs completely regular in the browser, it may work without any framework. This stimulates scalability and portability of applications. Designed models can be saved and restored. However, Figure 4 shows an example of using GoJS to create an object on the Canvas. Code on the left is to declare a node with a rounded rectangle shape with fill color properties and its caption. This node is drawn in a canvas called “canvasdiagram”. This node can be developed to include ports for input and output and link flows from this node to other nodes. A toolbox is created to define a plate of categories for the node instances is created, as will be shown, to build the entire tool. The designer is enabled to drag an instance of a node from the list of categories to build the wanted diagram. The illustration of this toolbox and its categories is presented next.

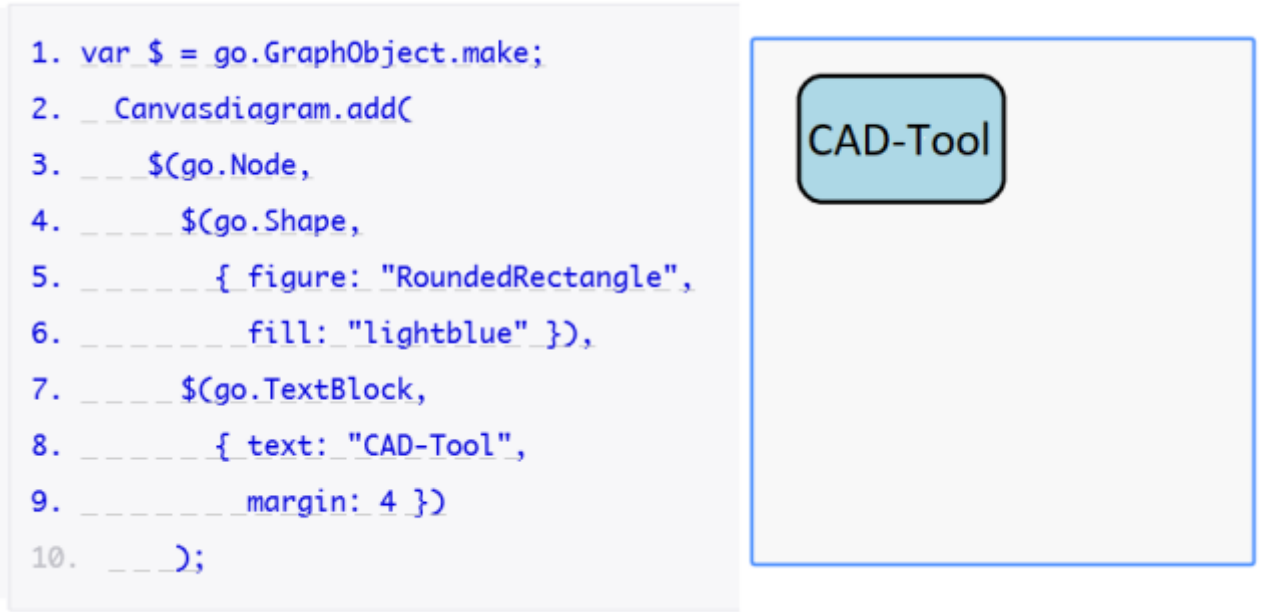

Figure 4. Sample of using GoJS library to create an object.

The first category required in the CAD-tool is the “Sensor”. A sensor is "A device which provides a usable output in response to a specified measured" [42]. In ARDEA, sensors are sampled at static rates. [43]. The symbol of a sensor is shown in Figure 5, the black rectangle at the right is the sensor outlet to the hardware processing element.

The second required category is the output devices or the “Actuator” category. This component interfaces a system to the surrounding environment. Output devices are usually transmitters or actuators. An actuator is a component of a machine that is responsible for moving or controlling a mechanism or system. The Actuator symbol is shown in Figure 6. Output devices are connected to processing elements on which the driver software modules reside.

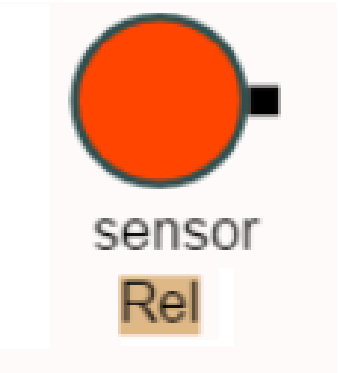

Figure 5. Sensor Symbol.

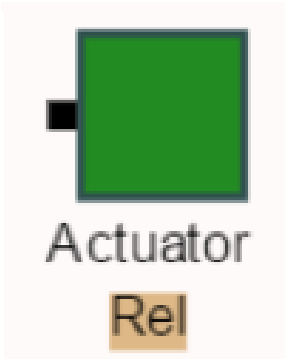

Figure 6. Actuator Symbol.

In ARDEA, the data produced by sensors is passed to a series of modules, the first module that the physical environment acquisition system faces is the Hardware Processing Element (HW PE). A hardware processing element could also be driving the actuator. If the processing elements are not directly interfering with the physical environment, then they will be Software Processing Elements (SW PE). However, each module performing its transformation before passing the data to the next module in the ARDEA pipeline is considered as a SW PE. Each module has a known execution time and resource requirements. Nothing distinguishes the HW PE from SW PE. Figure 7 shows the symbol used to represent the processing element.

Data Variable (DV) nodes represent the information passed between modules. A special type of data variables is Management Data Variables (MDV), only the system manager deals with this type of data for tracking or reconfiguration purposes. Figure 8 (a) shows the data variable symbol, while Figure 8 (b) shows the MDV symbol. Dependency Gates' nodes are the nodes that control the flow of data variables from the sources (sensors) of the pipelines toward outputs (Actuators) of the pipelines. Symbols of OR, AND, and XOR gates consequently are shown in Figure 9.

The complete set of ToolBox components are shown in Figure 10. The brief function for every part is shown next to every component. The Toolbox contains Comment facilities, Mux, and Demux, some components are added just for assigning them a function for future use in case of the desire to expand the functionality of the tool.

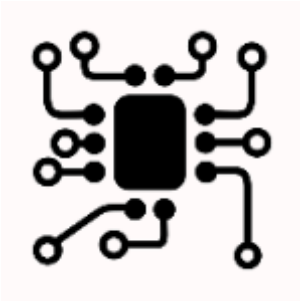

Figure 7. Processing Element.

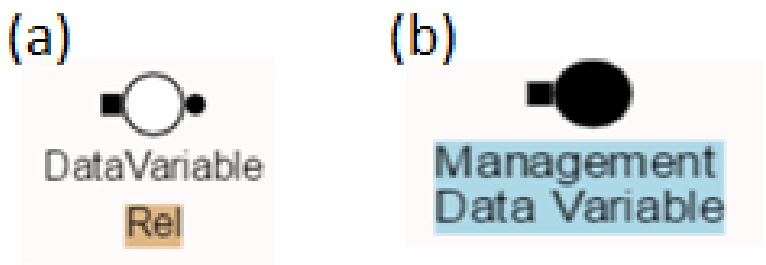

Figure 8. Data Variables (a) Regular Data Variable symbol. (b) Management Data Variable.

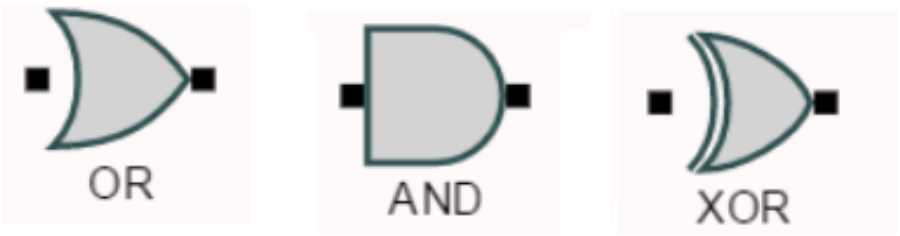

Figure 9. Dependency Dates (Gates) used.

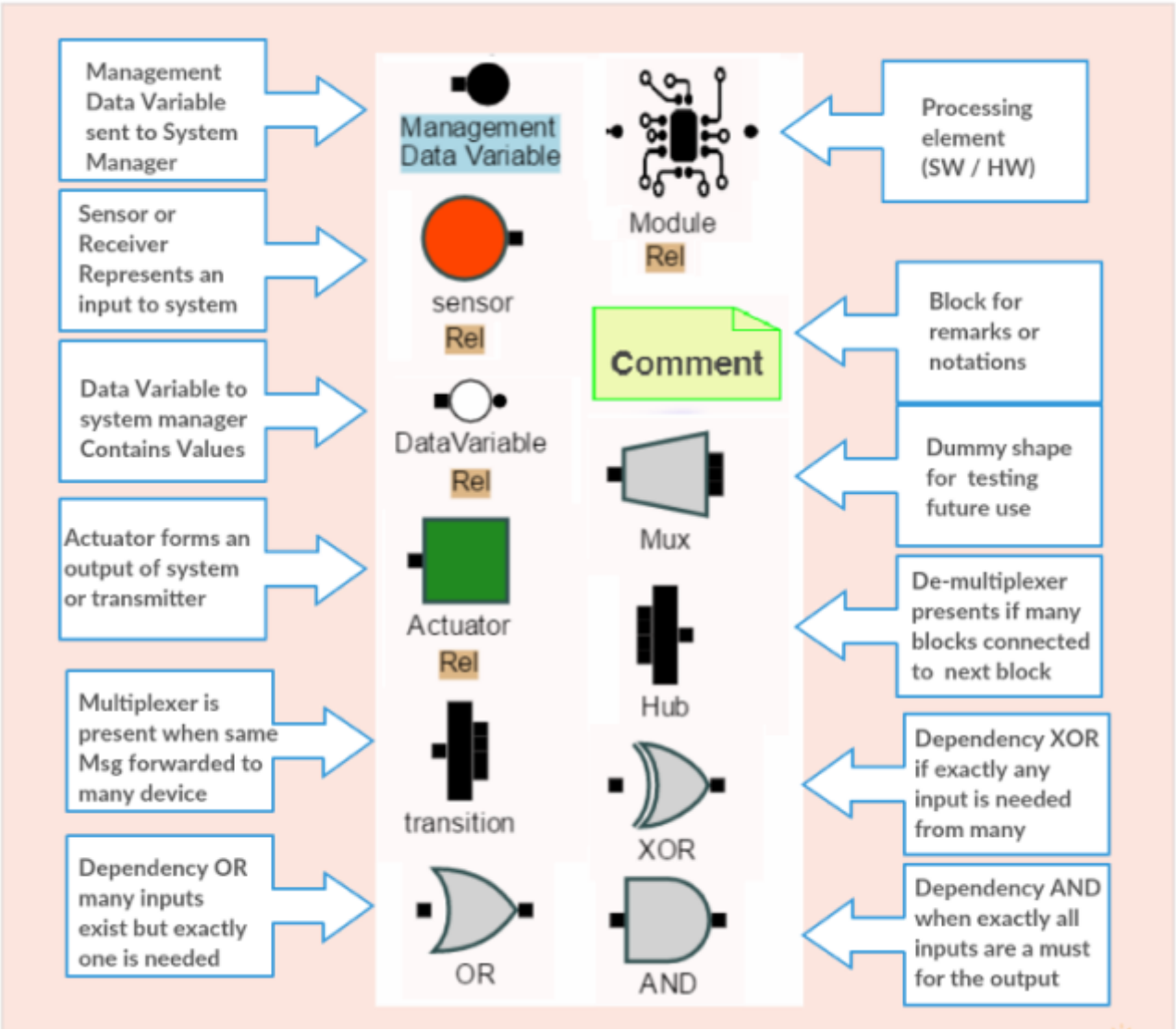

Figure 10. Developed Toolbox of the CAD-tool.

## 2) USING THE CAD-TOOL

Creating ARDEA DGs are developed using the proposed CAD-tool by specifying the flow of data and the dependencies among software and hardware modules that perform the required computations. Data variable nodes represent the information passed between modules. The data variable (or input) requirements of a software module are represented in the graphs using a set of dependency gates, allowing the specification of a module with several distinct input requirements. The software modules therefore form the basic degradable units of the architecture.

An example of using the tool to design a simple serial DG for DES blocks using the proposed CAD-tool is shown in Figure 11. The environment of design shows the tool built on a web browser. In the left side toolbox appear all DG nodes. On the workspace or the Canvas on the right there are five components that are serially connected, “sen1” is the input, which is connected to “Mod1”. “Mod1” processing element is a HW PE that samples the sensor data and works as a data-conditioning element. “DV1” forms an output DV that will be sent to the next PE. “Mod2” as the next hop PE receives this message and does the required processing before forwarding the message to its final destination “Act1”.

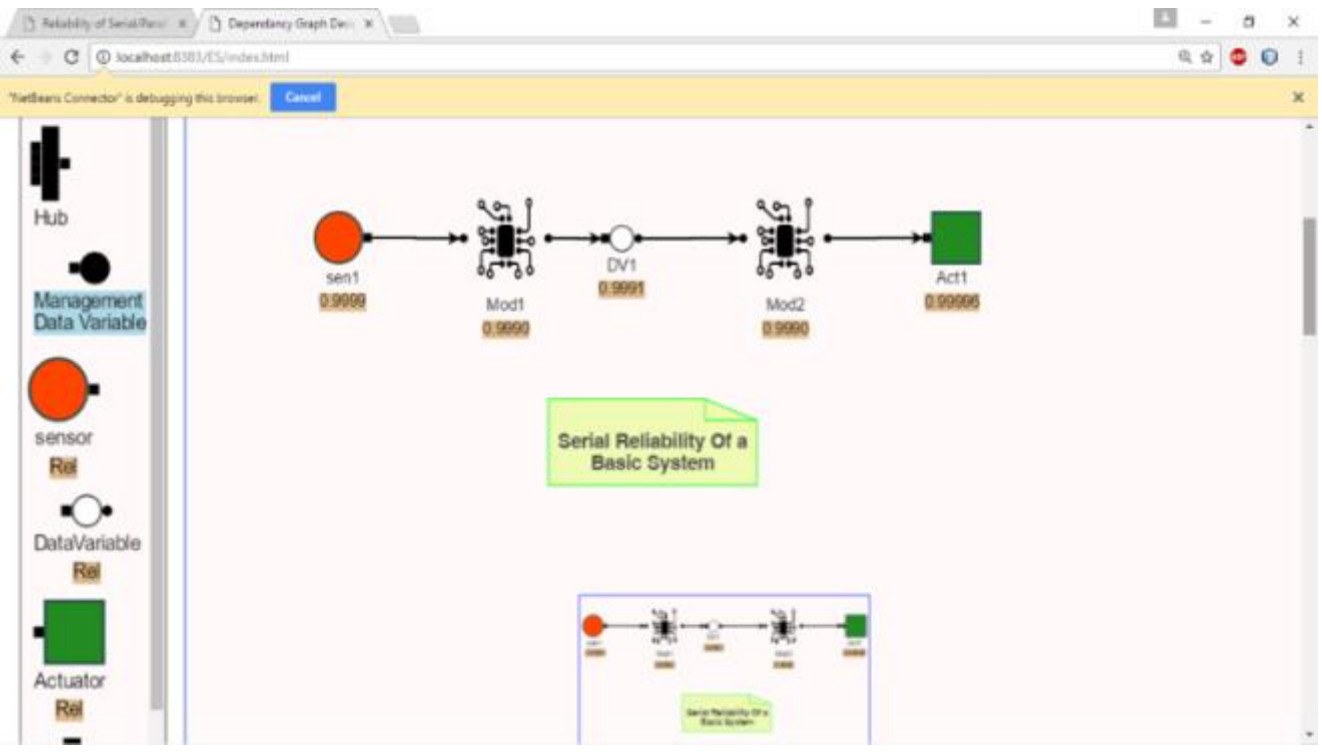

Figure 11. An example of a simple serial configuration.

In this example, there are no more than one option in this graph because it is a single serial path. The below box can be used to navigate the workspace when the user would like to view hidden nodes out of the current window scope. To enable the user to save the DGs’ settings in configuration files, this tool is armed with backup and restore capabilities. The CAD-tool will first bind the graph to a JavaScript Object Notation (JSON) formatted text file. This binding is in two directions, the text is transferable to a graph, and the graph can be converted to a text. A JSON file generated from saving the previous serial configuration is shown in Figure 12.

As a recapitulation about JSON file, JSON is a general-purpose data interchange text format defined as a subset of JavaScript object syntax. JSON is built on two structures: A collection of name/ pairs, which is realized as a record, an object, hash table, keyed list, or an array. The second structure is values in an ordered list, in most languages this is recognized as an array, vector, list, or sequence. JSON is a universal data structure that all the modern programming languages support, which makes the use of such interchangeable format with programming languages extremely feasible.

However, when the designer saves the graph using the tool, it will generate the configuration file shown in Figure 12. The JSON file describes the properties of the models’ nodes as a set of arrays. Every node has its own key and other specifications, such as node location, caption, and other specifications. Another set of arrays follows to describe the links map between nodes. However, when the designer saves the configuration, the tool will automatically add both “Start” and “End” nodes to the design. Start and End categories are used later as delimiters in processing the graph. When saving the configuration file, the Start and End nodes will automatically be saved. The user can keep a repository of previous files in his own folder and just copy the entire configuration code and paste it into the tool, and the graph will be generated. If the designer would like to change or amend the graph, he can back-up his new changes on his PC. The size of the configuration file is relatively small; the previous file requires only 1,327 bytes.

More advanced examples taken mainly from ARDEA framework [2] are implemented next. Figure 13(a) shows a three pipelined DG to use N-version redundancy for fault tolerance. A voter module handles transient faults and chooses the majority. Two out of the three input sensors’ readings (pressure sensor) have to correct to produce a correct altitude estimate for the Unmanned Aerial Vehicle (UAV).

Another example is demonstrated in Figure 14. Figure 14 (b) is the implementation of Figure 14 (a) using the developed CAD-tool. This implementation shows the degree of similarity and proficiency for the tool to enable the designer to accurately design the required diagrams.

Moreover, the example in Figure 15(a) relates a flight controller that uses two sources for tracing information, GPS receiver and magnetometer. Dependency gate XOR is used to specify that only one input is required at a moment. The management data variable (MDV) is sent to the manger for tracking and reconfiguration purposes in a case of failure. Again, the tool is used in Figure 15(b) to support such examples. Figure 16 illustrates another example utilizing more Dependency gates.

```
"class": "go.GraphLinksModel",
"linkFromPortIdProperty": "fromPort",
"linkToPortIdProperty": "toPort",
"nodeDataArray": [{
        "category": "sensor",
        "key": "sensor 1",
        "loc": "-790.6837592529297 -478.21550625",
        "name": "sen1",
        "Rel": "0.9999"},
        {"category": "Module",
        "key": "Module 1",
        "loc": "-668.1232749323846 -498.7627312620751",
        "name": "Mod1",
        "Rel": "0.9990"},
        {"category": "DV",
        "key": "DV 1",
        "loc": "-551.4346749323846 -467.15956256761564",
        "name": "DV1",
        "Rel": "0.9991"},
        {"category": "actuator",
        "key": "Actuator 1",
        "loc": "-279.16127493238446 -478.0991188176156",
        "name": "Act1",
        "Rel": "0.99996"},
        {"category": "Module",
        "key": "Module 3",
        "loc": "-433.53056868238446 -498.762731262075",
        "name": "Mod2",
        "Rel": "0.9990"},
        {"category": "label",
        "key": "label",
        "loc": "-593.9773936823844 -331.428036727202",
        "text": "Serial Reliability Of a Basic System"}],
"linkDataArray": [{
        "from": "sensor 1",
        "to": "Module 1",
        "fromPort": "out",
        "toPort": "in"},
        {"from": "Module 1",
        "to": "DV 1",
        "fromPort": "out",
        "toPort": "in"},
        {"from": "DV 1",
        "to": "Module 3",
        "fromPort": "out",
        "toPort": "in"},
        {"from": "Module 3",
        "to": "Actuator 1",
        "fromPort": "out",
        "toPort": "in"}
 ]}
```

Figure 12. JSON formatted text configuration-file for serial path shown in Figure 11.

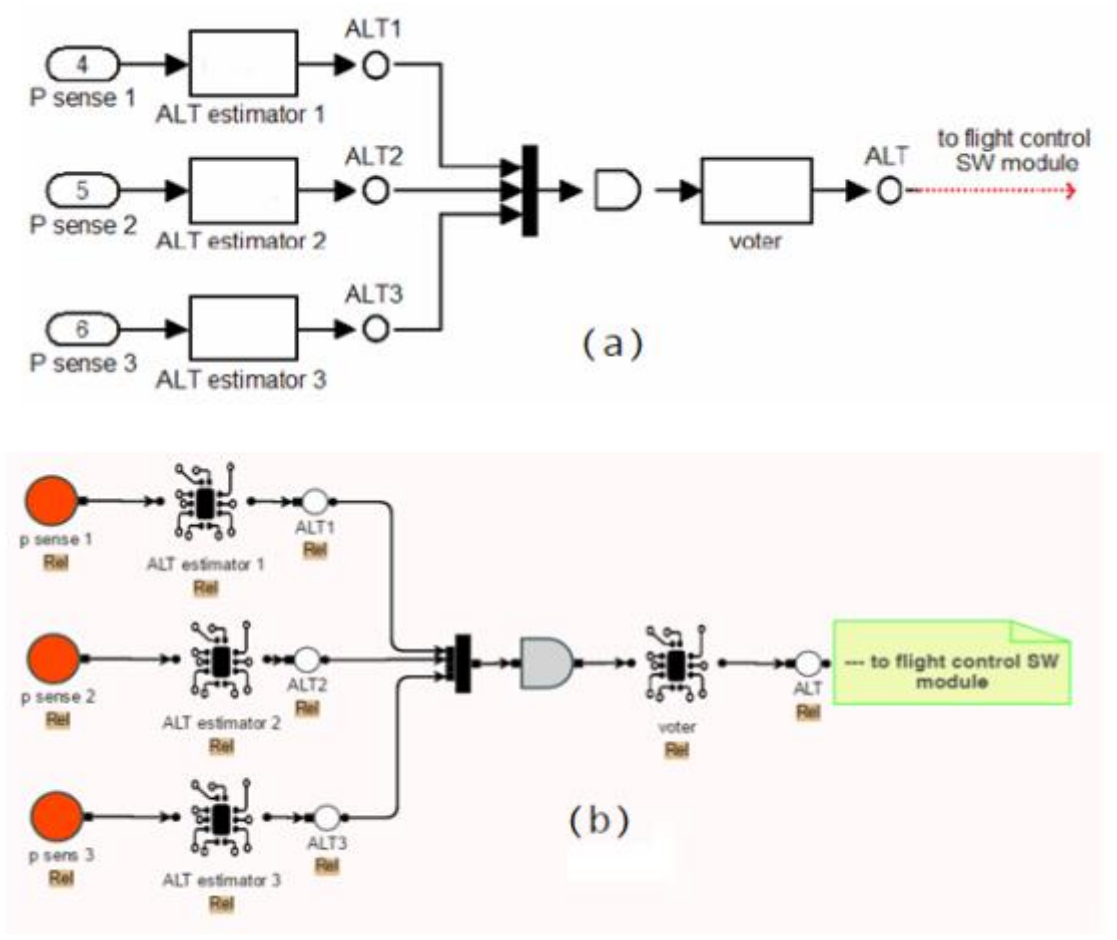

Figure 13. Triple Redundant Input Devices with Voter (a) DG created manually (b) DG created by the CAD-tool.

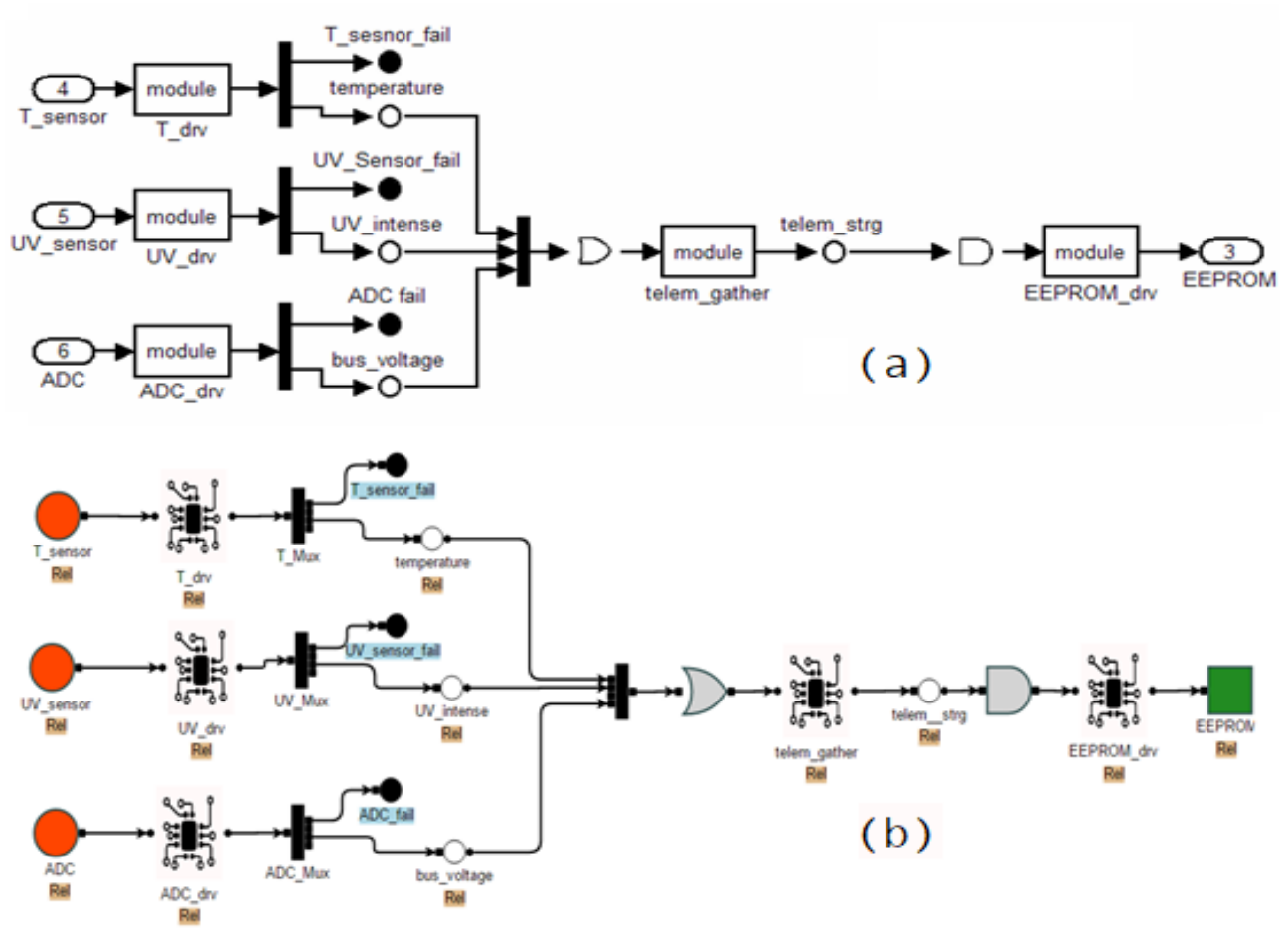

Figure 14. (a) DG created manually (b) DG created by the CAD-tool.

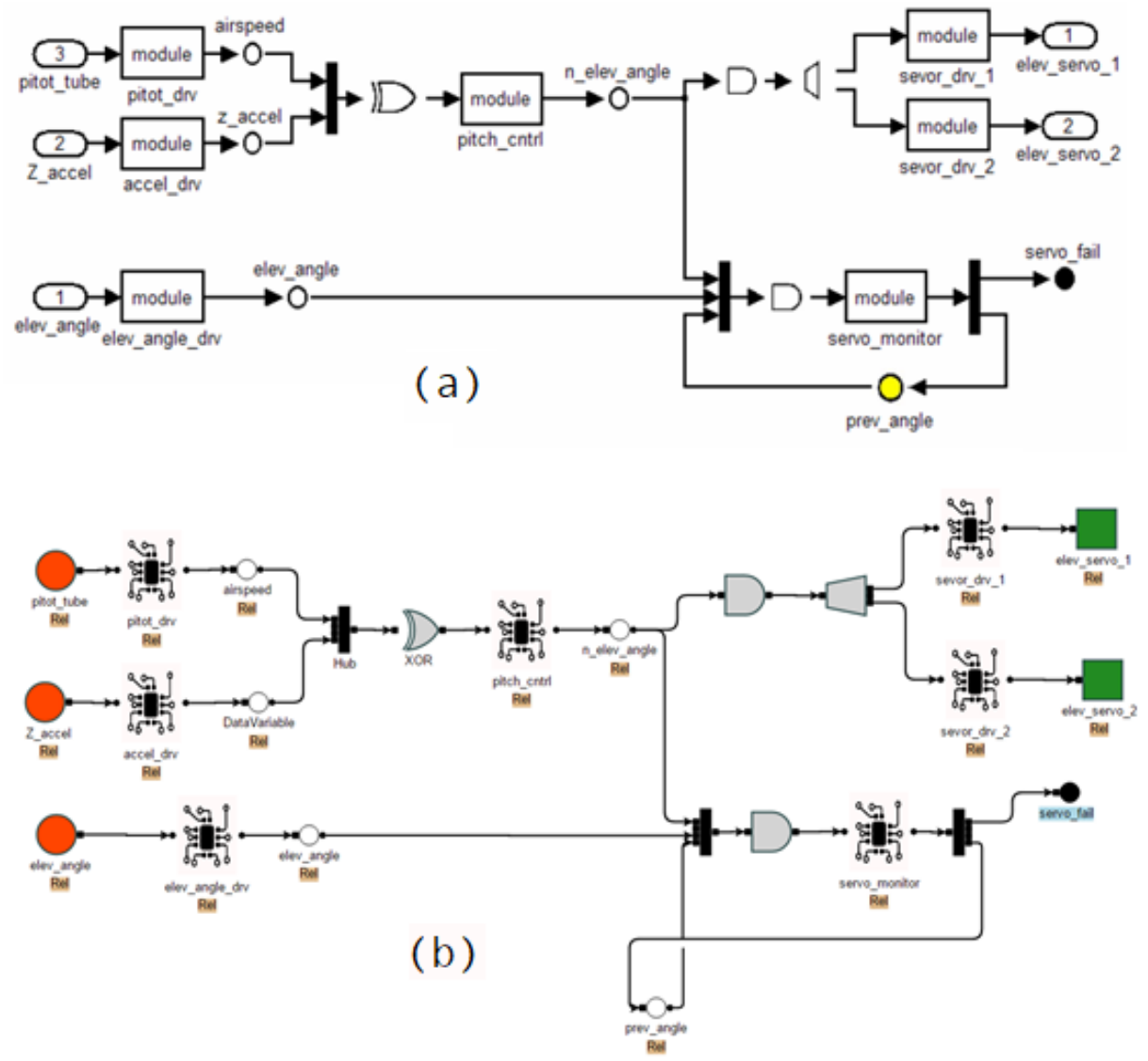

Figure 15. (a) DG created manually (b) DG created by the CAD-tool.

### 3) ENGINE OF THE CAD-TOOL

To highlight more potentials for the developed tool, parallel DGs of a couple pipelines are considered. Figure 17 shows the DG in case of two parallel path. Reliability variables of elements can be filled directly under items on the tool, some elements do not need to be given any variable, because their only role is to describe and control the flow of data variables. Start and end nodes will be added to the graph automatically when saving the DG as shown in Figure 17.

An engine used to parse DGs and extract the pipelines. A pipeline expresses the chain of conditions required to ensure flow of data variables from data sources to their destinations. Or from the sensors on the left hand side until reaching successfully to the actuators on the right hand side. The flow diagram for DG parsing engine is shown in Figure 18. The result of processing the diagram of the parallel/serial example in Figure 17 is shown in Figure 19 directly under the graph. There are two pipelines produced:  the first sequence is (sensorOne, FirstModule, FirstDataVariable, ThirdModule, and Output), and the second sequence is (sensorTwo, SecondModule, SecondDataVariable, ThirdModule and output).

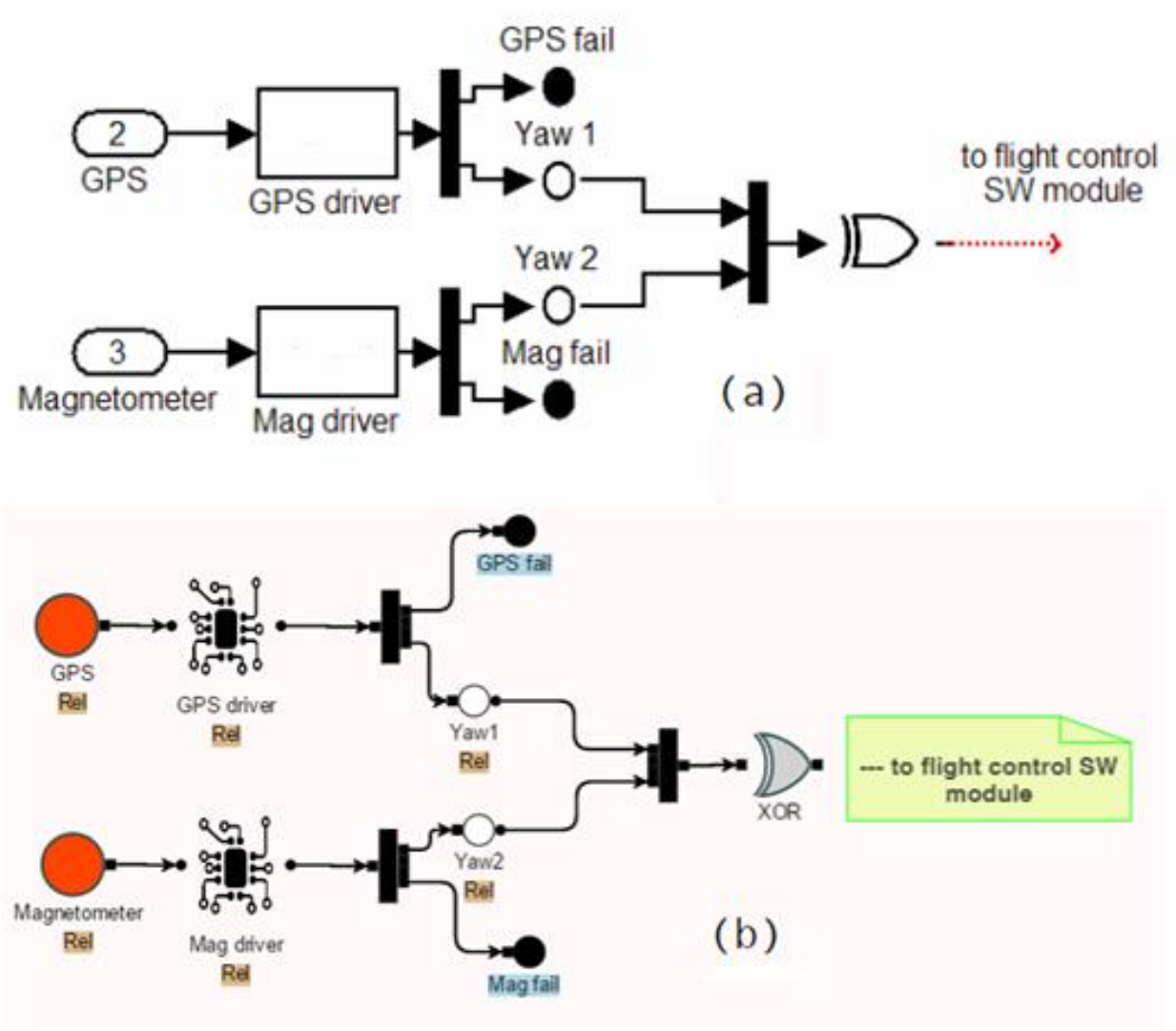

Figure 16. (a) DG created manually (b) DG created by the CAD-tool.

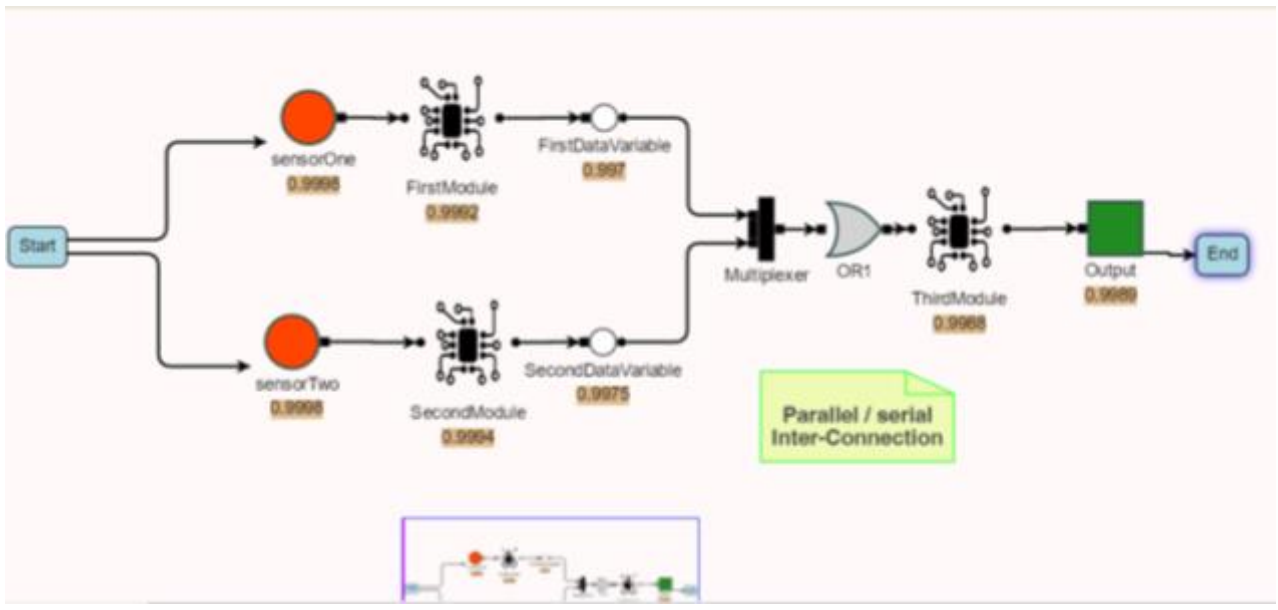

Figure 17. Adding start and end nodes to the connections.

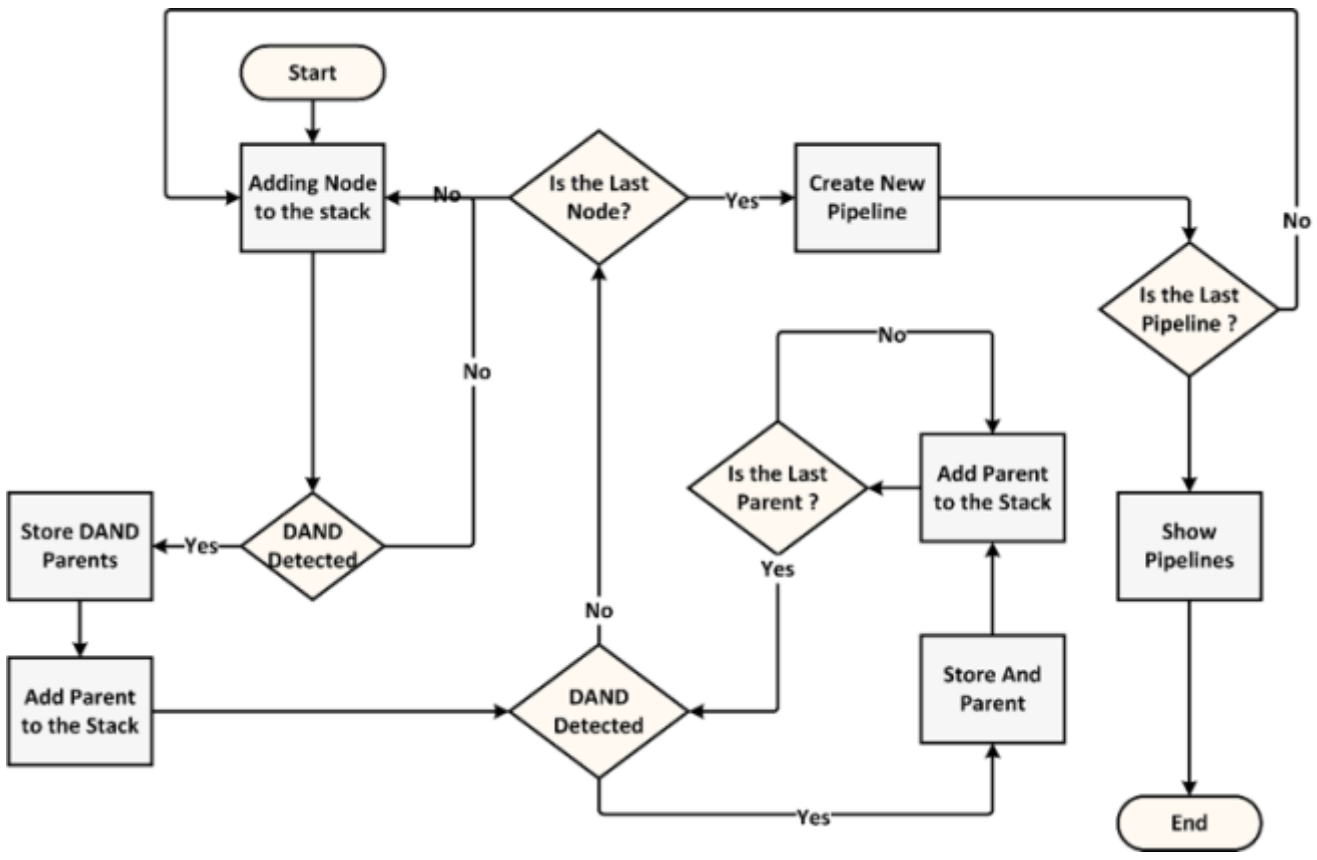

Figure 18. DGs parsing engine flow diagram.

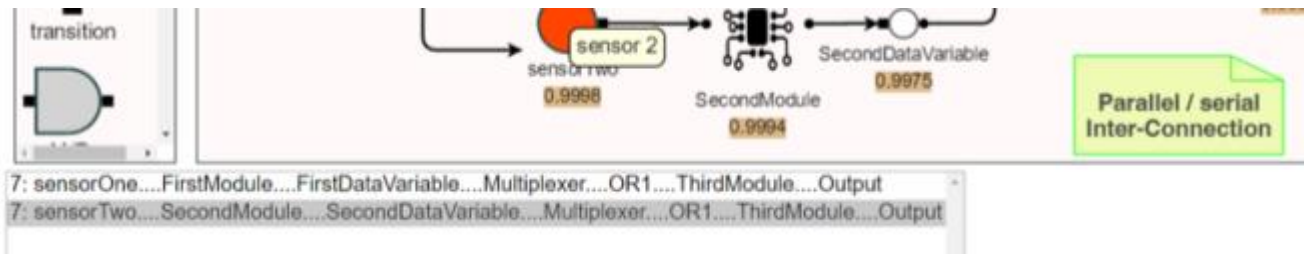

Figure 19. Pipelines after parsing DGs of Figure 18.

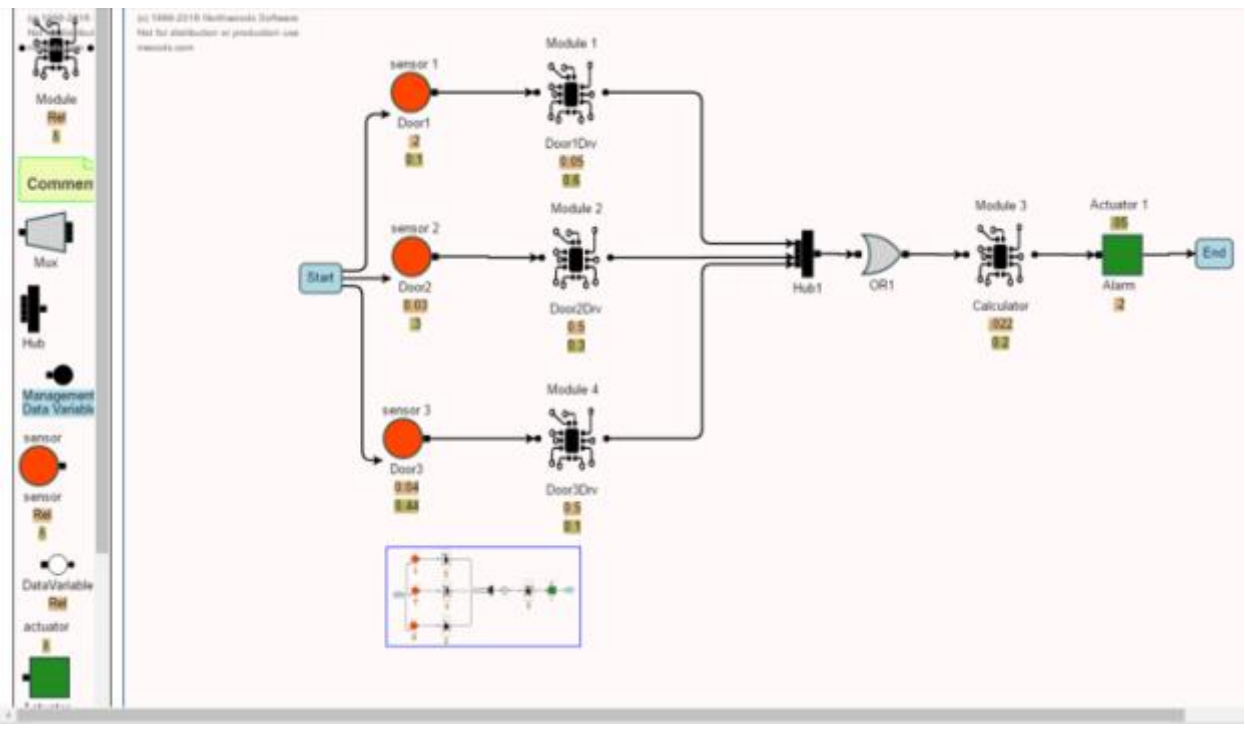

Figure 20. Three redundant systems DG created.

The engine searches the end node and adds nodes into a stack until the start node is reached to create a full pipeline, then the engine pushes the pipeline to a collection of founded pipelines. If the parser finds a dependency AND gate or DAND to simplify referring to it, then it will serialize all branches of the parents of the DAND gate and complete the path. Other gates and nodes are handled using JavaScript programming. Figure 21 shows the parsing of DGs as represented in Figure 18. The result of parsing is two possible pipelines extracted to keep the output node functioning.

To put all together, the DGs shown in Figure 20 parsed, the results appear under the DG. They are three pipelines. When the user selects the three pipelines sequentially, every pipeline will reveal its details.

Figure 21(a1) shows that the designer has selected the first option on the list, and he would like to investigate the first reliability rank pipeline for verbose study. Figure 21(a2) automatically will appear and give the details of the path. Figure 21(a3) represents the reliability curve over the time for the first pipeline. The user can slide the pointer over the curve to discover the reliability value of the pipeline in function of time. Figures 21(b1), (b2), (b3) specify the corresponding findings for the second rank pipeline, and finally, Figures 21 (c1), (c2), (c3) are for the third pipeline.

It is superficial from Figure 22 that the reliability of the upper curve for the first pipeline is the best among others. However, if Door1Drv processing element or its acquisition element “Door1” sensor malfunctions, then the next option will choose “Door2” as source of information, until pipeline one is restored. The last alternative will be the third pipeline.

Every processing element in the system is given an ID number, ID is in hexadecimal format. This ID will be used later on to track the status of the processing element, and should contains one bit set only. Valid IDs set like 0x0001, 0x0002, 0x0004, 0x0040, and so on. A Logical OR operation between the IDs gives a hexadecimal number, this number reflects the status of the processing elements within this path. If the status number of a pipeline is 0x0E, then the working processing elements IDs are: 0x02, 0x04, 0x08, and so on.

The CAD-tool exports an array of numbers that forms the set of expected working elements in every pipeline. This array will be formatted in JSON data file, this JSON file typically contains text such as: “{[8476, 4007, 8478, 8479, 8475]}”, and ready to be exported outside the tool. However, one role for the system manager processor is to rely on this JSON file, as to be explained in the tool implementation for the case study. No confusion should occur between configuration JSON file for saving and restoring the DGs on the tool with the JSON file holding the array of reliable pipelines’ options which are sorted in a descending order, and will be exported outside the tool and inserted into the System managers of DES.

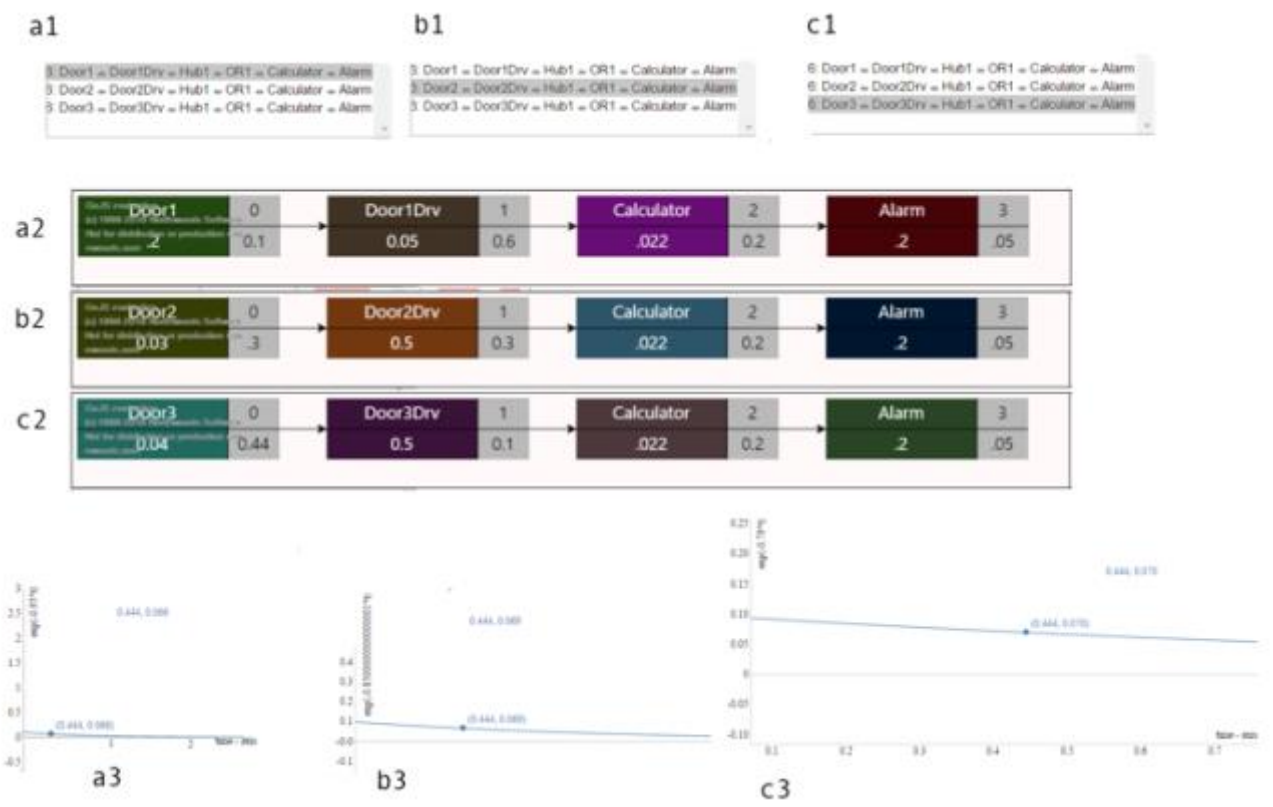

Figure 21. CAD-tool results of implementing triple redundancy.

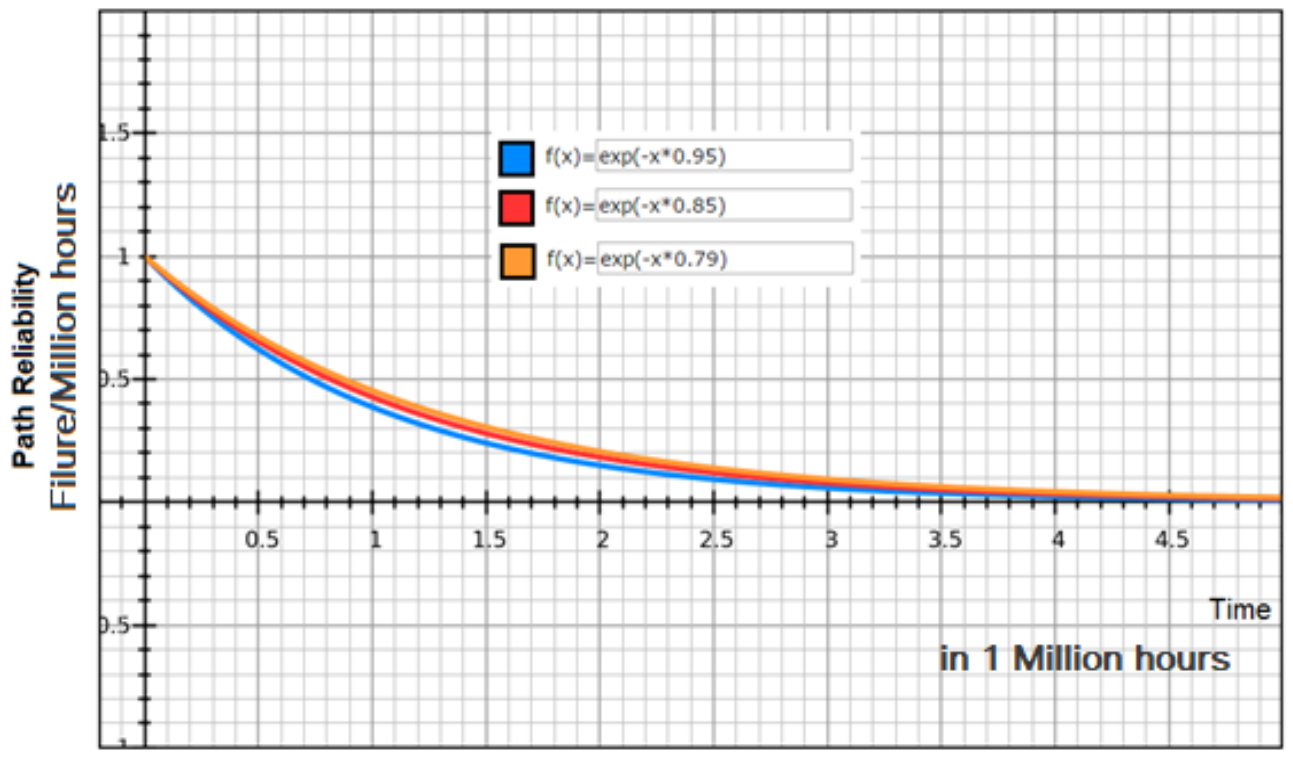

Figure 22. Comparing the reliability curves generated from the tool for the triple example.

## 5. EXPERIMENTAL SETUP, ANALYSIS AND RESULTS

### A. *Experimental Setup*

In order to implement the CAD-tool and use it in the case study, it is important to break the system down into the basic blocks to setup the required experiments. Instead of starting with the complex case study from the first time, the focus remains on the three parallel pipelines example, Figure 23, to realize the concept of the tool and use it on real-time DES. The case study comes after with some differences. In order to make the tool suite the applications, nodes of the tool are developed to give every module an ID, reliability, and failure rate variables. Experiments cover data file generation and file transfer process to plug it into the DES system.

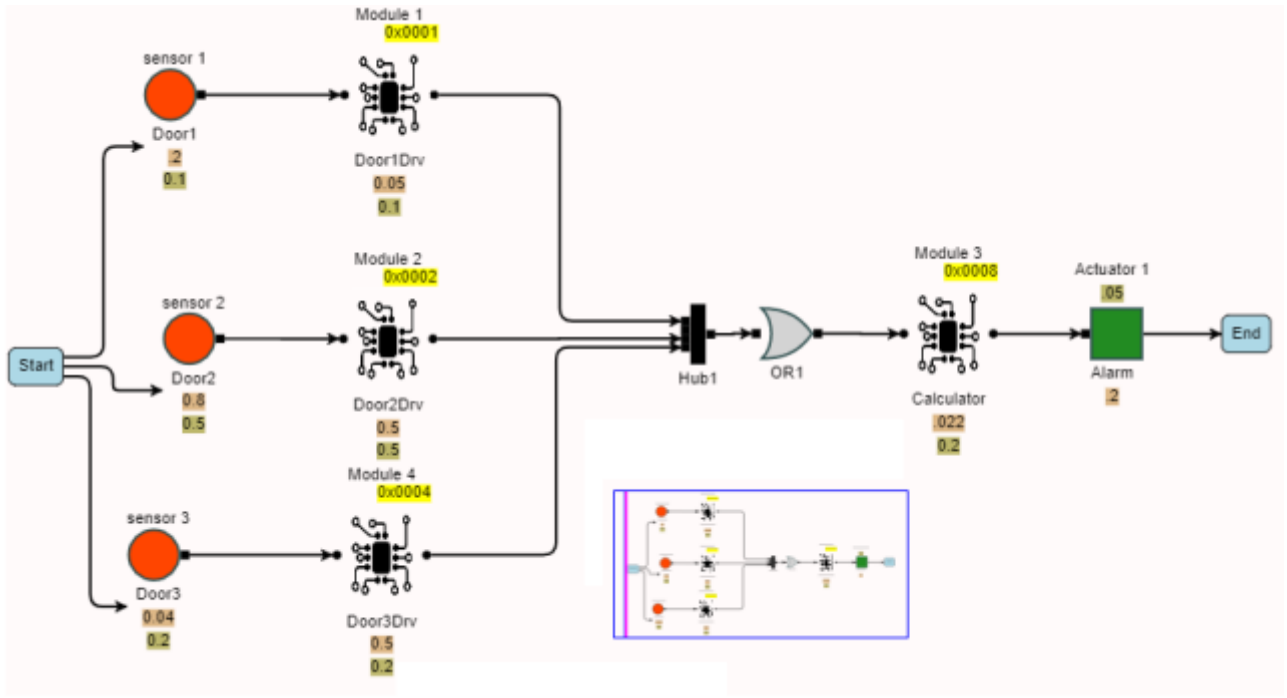

Figure 23. IDs assigned for the modules of the three parallel paths.

Every processing element is given an ID on a yellow space above its shape in the CAD-tool, this ID should match the processing element ID in application software. It is the responsibility of the designer to assign correct IDs on the tool and in implementation. The CAD-tool exports a JSON file that contains only text data, the exported data file for the three branches example contains the following string: {[9, 10, 12]}, the hexadecimal numbers are 0x09, 0x0A, 0x0E consequently. The JSON file contents are transferred to the system manager.

#### 1) PROCESSING ELEMENTS (PES)

The block diagram of a processing element (PE) is shown in Figure 24 from [2]. The lower portion of a PE is the physical hardware component. Above the hardware, a multitasking Real-Time Operating System (RTOS) lies at the bottommost of software layers, and application software modules are at the top. The set of software modules contains a local management software module which acts as the agent of the DES system manager on the local PE to schedule data forwarding among nodes. The processing element has hardware peripherals like ADC, PWM, and GPIO, etc. to facilitate communication with the outside world. A PE has hardware for networking facilities for communicating with other PEs and the manager of the DES. In this work, experiments use the MC9S12DG256 microcontroller, a member of the HCS12 family by NXP Semiconductors (www.nxp.com).

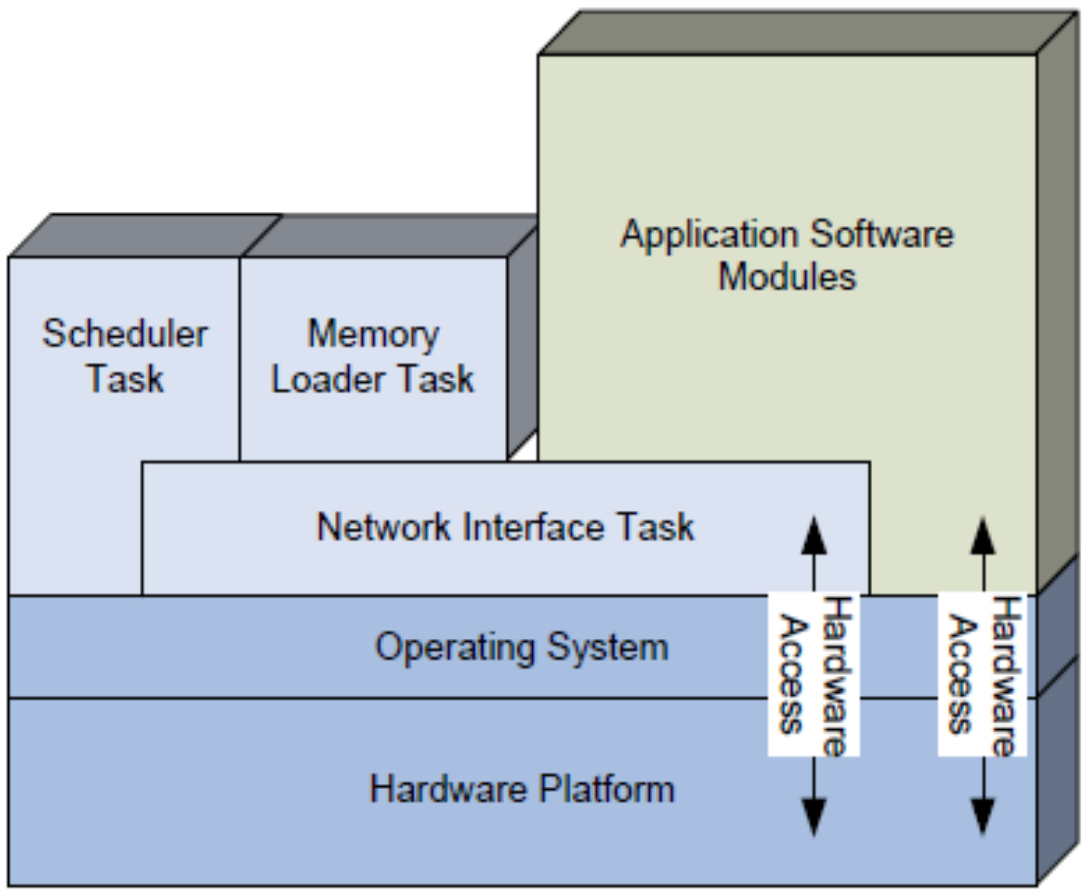

Figure 24. Hierarchy of Software on a PE from [2].

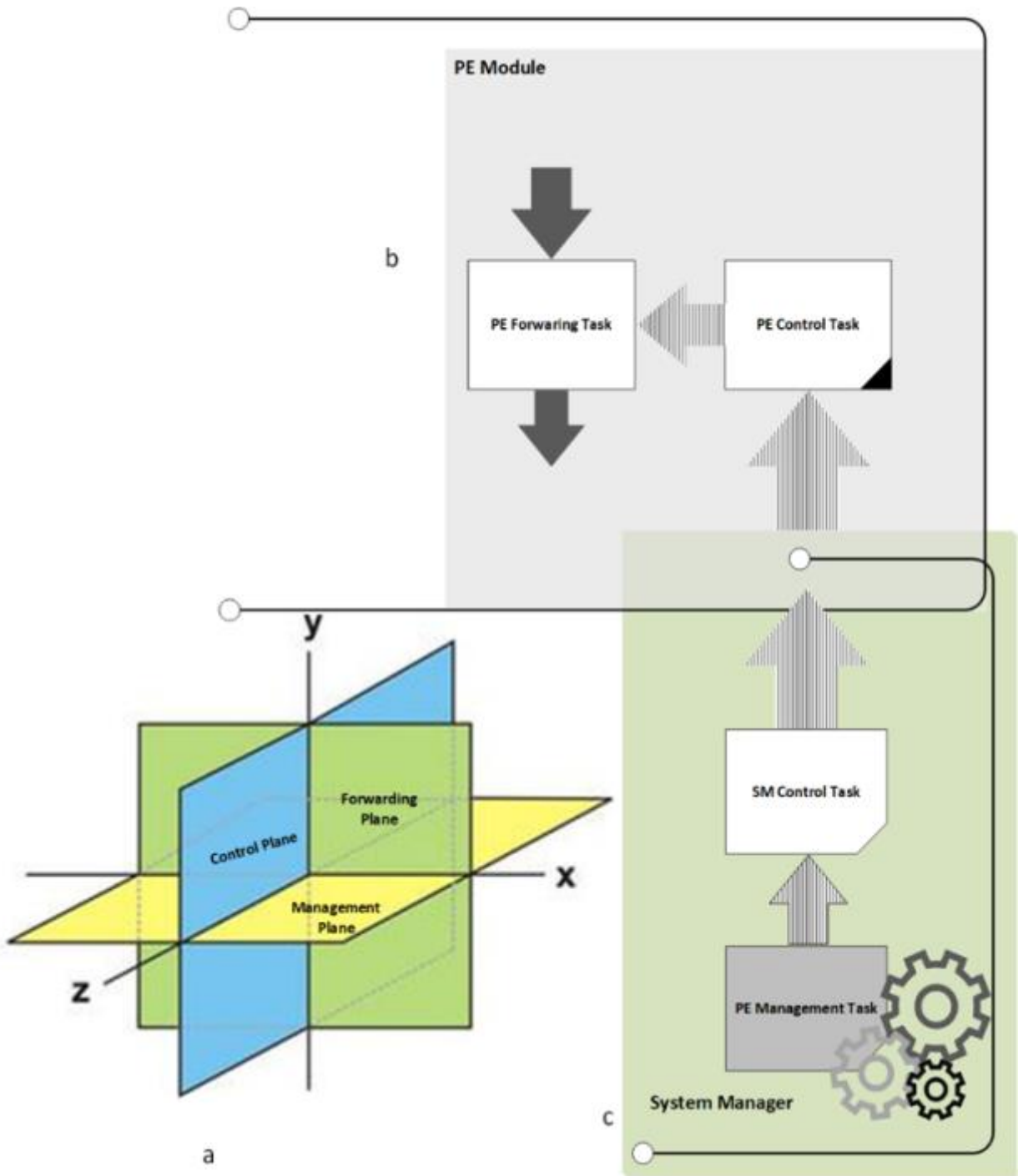

Figure 25. Conceptual planes for the tasks. (a) Planes interaction (b) A PE tasks. (c) System manager PE task.

The Real-time operating system (RTOS) used in the experiments is µC/OS-II from Micrium (https://www.micrium.com) [44]. In this work each of the many tasks of a PE can be attached to one of three conceptual planes of operation: first, Forwarding Plane, in which a frame transverse from an input to an output port. Second, Control Plane, which determines how frames should be forwarded. Finally, Management Plane, which determines the techniques of configuring the control plane. Networking applications by some vendors like cisco systems [45] uses this concept. However, the implementation of the mentioned planes on the PE software tasks can be applied by defining at least two tasks on every PE. One task to serve as the forwarding plane, and the second as the control plane. When the PE is a system manager a third task should be assigned to the management plane, the third task allows the designer to intervene with the system if he desires to change the working strategy.

Figure 25 (a) illustrates the concept. Figure 25 (b) shows the HW and SW PE modules plane, and Figure 25 (c) shows the system manager. A PE contains only two planes, one for forwarding and another for control. The forwarding task for a HW PE can be such as sampling sensor readings of speed or temperature and sending it to the SW PE estimator. On the other hand, the control task plays the role of an agent for a system manager. This agent controls the local forwarding task, by switching the forwarding process, or by tuning forwarding variables.

## 2) COMMUNICATION

For communication purposes, ARDEA framework recommends CAN network [46] for intercommunication among the DES PEs [2]. Many features strengthen CAN to be used as a communication protocol. It is low-cost, lightweight, and the transmission media is typically two twisted wires [47]. CAN is a CSMA/AMP (Carrier Sense Multiple Access/Arbitration by Message Priority) [48], multi-master serial bus standard [46]. A transmission can carry up to 8 bytes preceded by an identifier describes the meaning of the data not the message destination. Identifier of CAN standard format version 2.0A is 11-bit long, but the extended version 2.0B has a 29-bit identifier.

3) SYSTEM MANAGER

In case of a system manager, the agent task is to serve as a global controller task for the entire DES, it communicates with the agents on all PEs. When the DES is not very distributed, the resources needed for control and management are not so much, a single task can comprehend the management and control duties.

A system manager is a special PE with extra functionality for communicating with the DES FT designer. An extra task is needed for a system manager to allow contacting the CAD-tool for management. This management task works on receiving the JSON file generated by the tool and holding the wanted strategy of DES. This strategy forms the control plane strategy, which controls the forwarding plane. The system manager receives JSON data file via Serial Communication Interface (SCI). The designer can use the well-known applications for serial communications as HyperTerminal which is shown in Figure 26, but other options could be but not limited to: Putty, TeraTerm, RealTerm, or any other serial communication software. Some of those applications require the designer to copy the contents of the file and paste them into the software window.

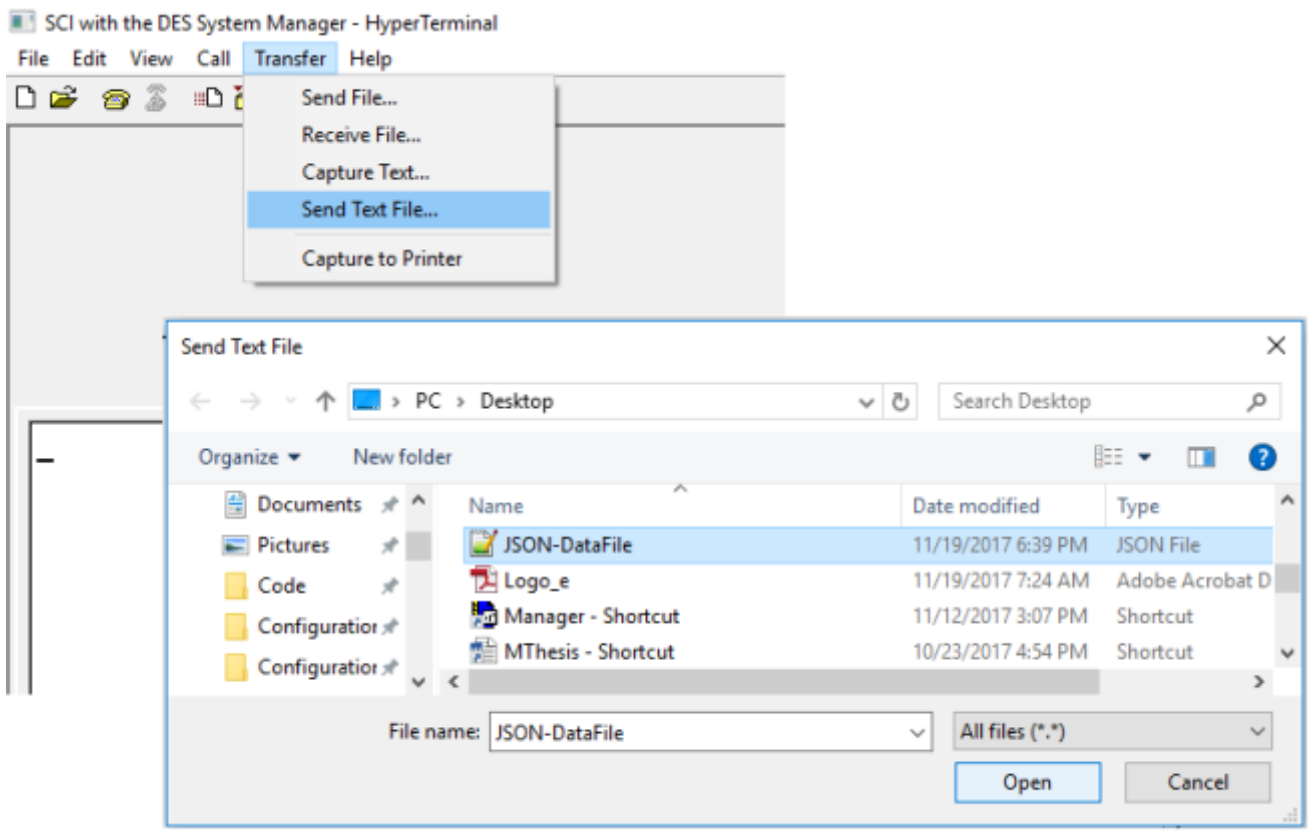

Figure 26. Transmit JSON data file to a system manager.

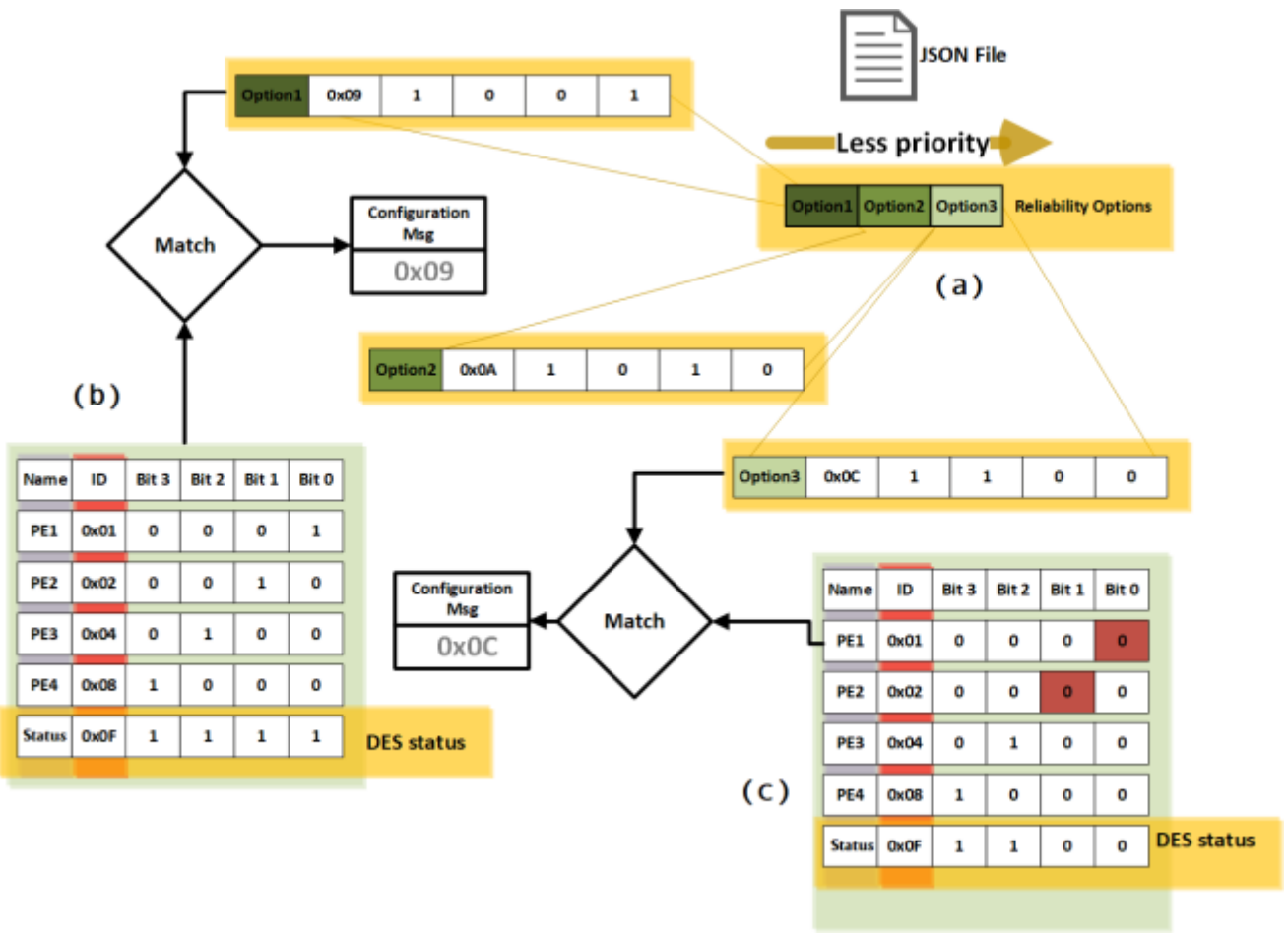

Figure 27. Selecting an option from Reliability options (a) Controller parser generates reliability options array. (b) DES status generated from PEs hello messages summation. (c) Comparison process triggers a new configuration message for the DES.

After that, the system manager parses the JSON formatted text file to convert the numbers in text format into integers to be used later within the control process. The parser is built around the lightweight JSMN parser [49]. JSMN is suitable for resource-limited ESs. After the parsing process, the parsed numbers are pushed into an array of numbers called “Reliability Options”. The reliability options array is considered a reference for the preparation of the new configuration plan. Figure 27 describes this preparation progress.

In Figure 27 (a) the controller keeps a copy of the reliability options in the Reliability Options array. Another record the control process on the system manager has to keep is the PEs status record. The controller process in the system manager has to get the status of all PEs in the DES. Every PE has to send a hello message every period of time, determined by the controller. If a PE does not send a hello message for a timeout period determined also by the controller, then the controller considers this PE as dead. This status register is of variable length according to the number of PEs in the DES, for the triple voter there are only four PEs, so the controller needs only four bits for the status register.

In Figure 27 (b), DES status register has 1111 binary number (0x0F), which means that all PEs are appropriately working. The controller will mask this status register with the options from the Reliability Options array, starting from the best reliable option which is option1. Option1 contains 1001 (0x09) which indicates that Option1 is matched if and only if the first alternative PE and the voter are working. Option1 is always selected as this condition stands.

In the scenario shown in Figure 27 (c), the first and second PEs breakdown, they will send a hello messages containing zeros instead of sending their IDs. The status register then will reset flags reserved for both PE1 and PE2, this gives a value of 1100 (0x0C) to the status register. Basically, this value is a minimum requirement for matching Option3, and actually this is the last resort in this reliability array for keeping the system up.

If a hello message is not received either due to a failure in a PE or because of a communication problem, then the controller process will wait an aging time period before clearing the PE specified bit and consider it failed. Despite of that the used aging time algorithm is unsophisticated, but the developer can vary this timeout to adjust the system firmness with absence of hello messages. Figure 28 illustrates the process of generating reliability options, and how does a system manager rely on them to control the DES.

### *B. Case Study: Fault- Tolerant Anti-lock Braking System (ABS)*

An anti-lock braking system (ABS) is taken as an example of safety-critical DES applications. ABS offers enhanced vehicle control and reduces the distance of vehicle stopping. It prevents the lock wheels or suddenly stop whenever a hard braking is applied. On snow-covered surfaces, ABS can significantly decrease stopping distance in an emergency, and improve steering control [50, 51].

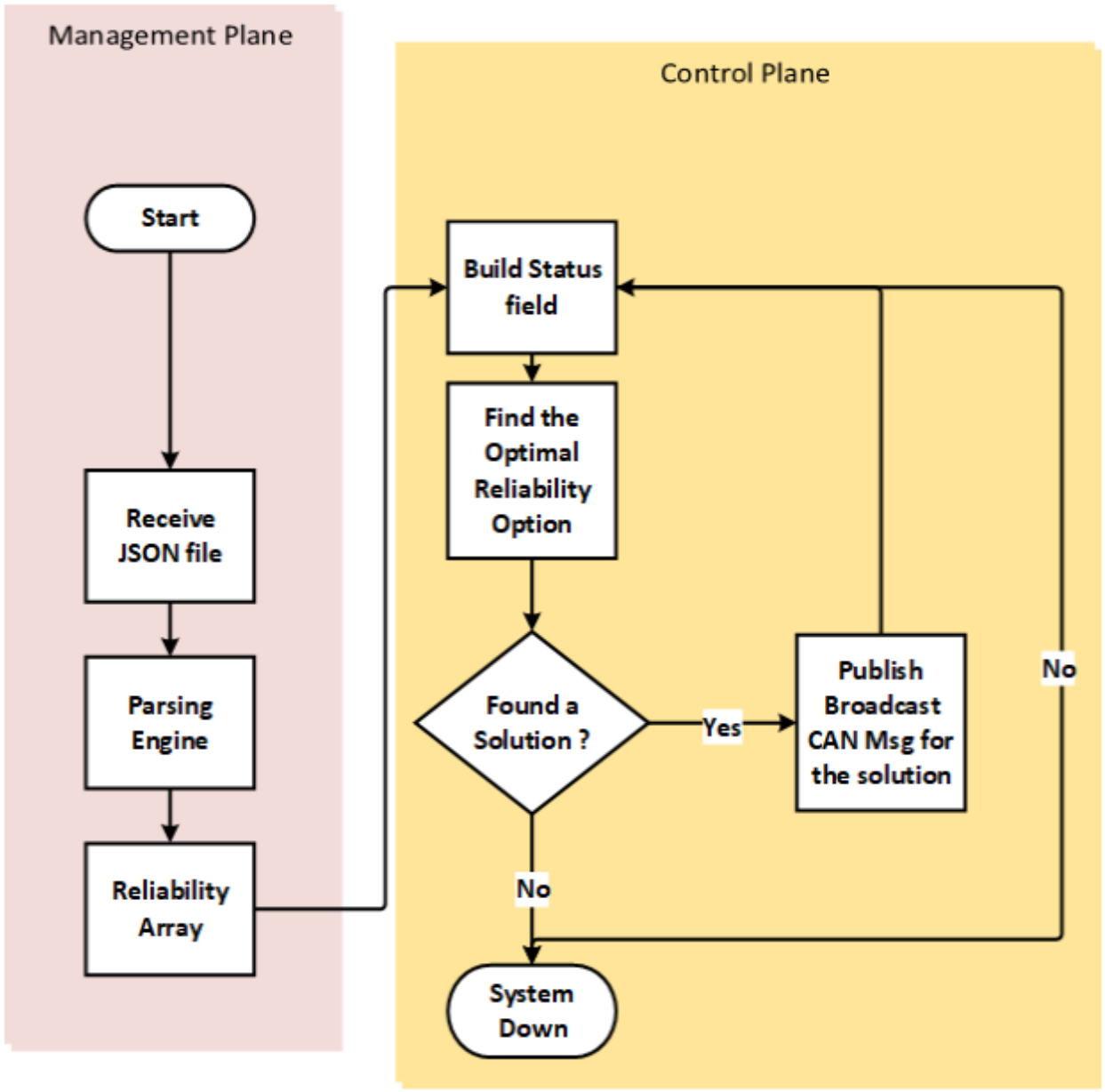

Figure 28. Controller Engine for a system manager.

ABS equipment is also used to implement a Traction Control System (TCS) on acceleration of a vehicle. When accelerating the tire loses traction, the ABS can detect the state and take appropriate action to restore the traction. Therefore, ABS focuses on the stopping and regulates braking of the vehicle, but TCS stops the wheels from slipping when they are accelerating by permitting just enough throttle application, and allows balance of traction and throttle.

ABS typically includes an Electronic Control Unit (ECU), speed sensors, and hydraulic valves. The ECU monitors the speed of the rotating wheels, if a drop of speed for the wheel is detected, then the ECU orders the actuator to decrease hydraulic pressure on the wheel, decreasing delivered pressure means decreasing the braking force which allows the wheel to go faster. In contrast, if ECU senses the wheel is faster than others are, it increases the braking power by increasing hydraulic pressure to the wheel, reducing the wheel speed [52, 53].

ABS uses different schemes according to brakes in use, number of speed sensors, and number of individually controlled valves [52]. Table 1 shows some ABS types. ABS of Type (1) contains a speed sensor on every wheel, and a separate valve for every wheel. In Type (2), the same arrangement of speed sensors, but with a common valve for both rear wheels, so number of channels changed to three instead of four. Type (3) ABS contains three channels for actuators and three speed sensors, two for front wheels, and a sensor for a differential. Differential is frequently used instead/with the back sensors, this is because when a vehicle rotates, outer wheels rotate in more speed than internal wheels. Either the ABS controller has to be programmed to tolerate this difference, or a differential speed sensor can be used instead.

TABLE 01

EXAMPLES FOR ABS TYPES

| *Type* | *Number of Channels (valves)* | *Number of Sensors* |
|---|---|---|
| 1 | 4 | 4 |
| 2 | 3 | 4 |
| 3 | 3 | 3 |
| 4 | 2 | 4 |
| 5 | 1 | 1 |

ABS Type (4) has two channels, one for rear wheels and one for front as a pair. However, type (5) has only one channel for only one sensor on the rear axle. Figure 29 from [52] illustrates the different Types presented in Table 1. Light connections indicate speed sensors, the gray heavy lines highlight the hydraulic channels, and the box on the rear axle is the differential.

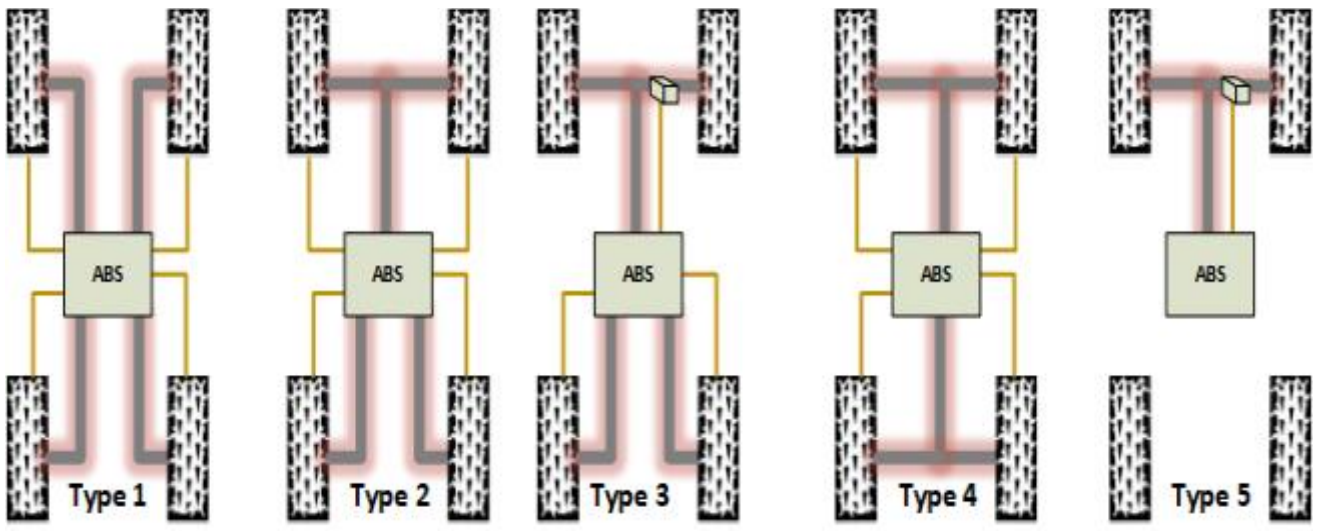

Figure 29. ABS types presented in Table 1 from [52].

ABS malfunction causes are numerous; it is often sensors' problems when metal shavings or debris contaminating sensors. The sensors connections are other common causes. Modules failure is less common, but also possible. The ABS is disabled if a fault is detected in any part of the ABS until the fault is cleared [52]. The work in this research equips the ABS system with self-healing capabilities through graceful degradation of the system in presence of faults.

The CAD-tool is used to design a Reconfiguration-Based Fault-Tolerant ABS architecture dependency graph based on what was presented by Sababha and Alqudah [52]. Figure 30 taken from [52] represents the DG of the proposed Fault-Tolerant ABS by Sababha and Alqudah [52]. In this DG, the rear right speed sensor (RR_Speed_sensor) is used to be a Hall Effect sensor where the number of pulses per second reflect the wheel speed. These pulses might be reshaped, regenerated, or encoded by the HW PE (RR_Speed_Est), output of this PE will be the data variable (RR_Speed). Because this pipeline faces an AND, that means that another branch is needed to be available for this path to be valid. The second needed branch of Rear Left speed sensor is (RL_speed_sensor), (RL_Speed_Est), and (RL_speed). Next SW PE (R_Speed_Diff_Est) gives R_Speed_Diff data variable which resulted of processing of the two data variables from RR_Speed_Est and RL_Speed_Est. In this case, it gives the difference between the two rear right and rear left sensors. A branch again is met after R_Speed_Diff and connects to the OR gate and passes to the AND, the second branch of this AND is taken from the Brake_Padel_Pos DV taken from the estimation PE brake_Padel_drv that estimates the position of the brake_padel via its sensor brake_Padel. The ABS_Control_Drv PE drives the ABS hydraulic pumps and valves according to the data variables arrived.

This pipeline forms an alternative for the ABS to keep working. Figure 31 shows all possible sequences that keep the ABS system resilient. Figure 32 illustrates how every solution comes, and its reflection on selecting the suitable sequence. In Figure 32 (1A), the rear set was selected, then a configuration message is sent to enable the highlighted sequence in Figure 32 (1B). The second pipeline is shown in Figure 32 (2A) with the necessary sequence Figure 32 (2B), and so on for other pipelines. Figure 32 (3A) and Figure 32 (3B) are for the third pipeline, and Figure 32 (4A), Figure 32 (4B) are for the fourth.

Regarding Figure 32 (5A) and the required sequence in Figure 32 (5B), it is a special case, because this path is not an alternative for the ABS system, but the system could be configured to prepare the TCS system to be a recovery module, this backup operation occurs when the ABS_driver module goes down, then the TCS system pipeline may be converted to a new working pipeline for the ABS.

The CAD-tool was able to find the paths and their relative Reliability Options array. When a user selects a path from the list located under the DG, the extracted path appears individually to focus on its calculations and to show its structure. This enables the user to review reliability variables. Simultaneously, the user can get associated calculations for this path. The entire Path reliability results, and the total failure rate are available. The result of the path reliability appears as a fixed number for a selected pipeline shown in Figure 33. But, the tool grants a reliability graph for the pipeline under study. The pointer can be moved over the curve to see the reliability over the time progress as shown at bottom of Figure 33.

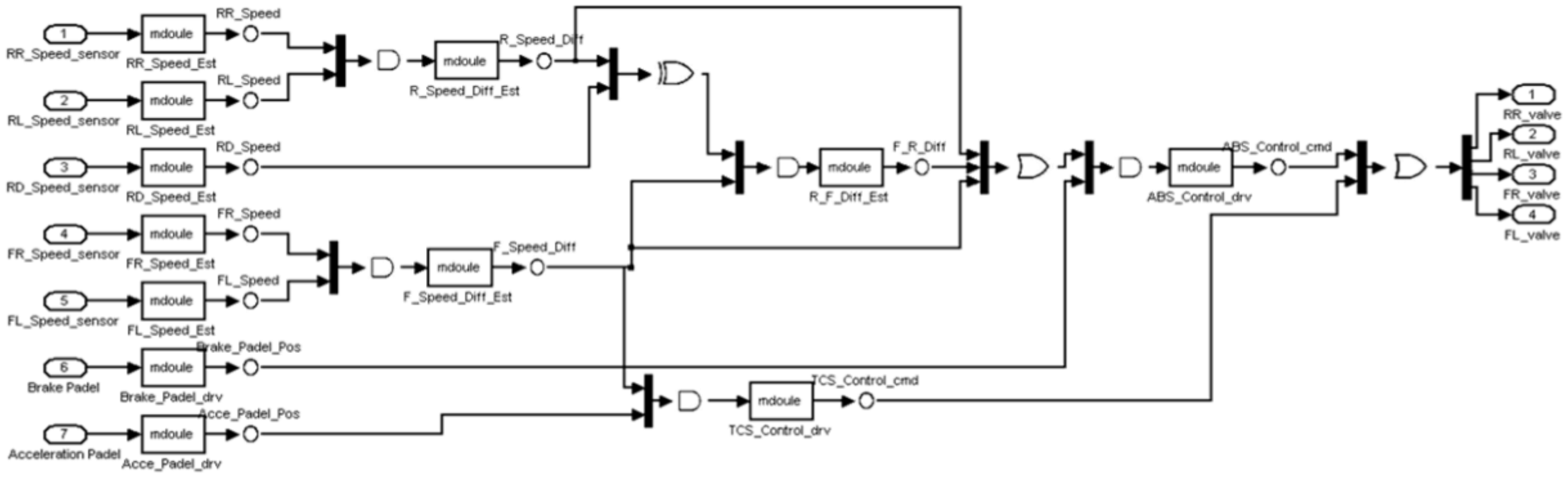

Figure 30. ABS architecture dependency graph from [52].

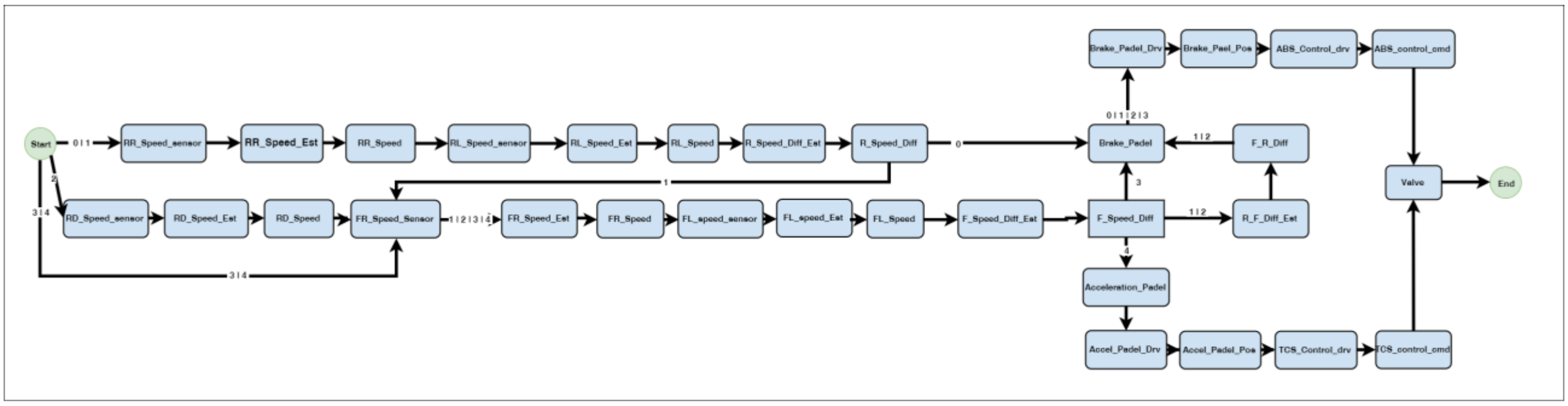

Figure 31. Possible paths for ABS.

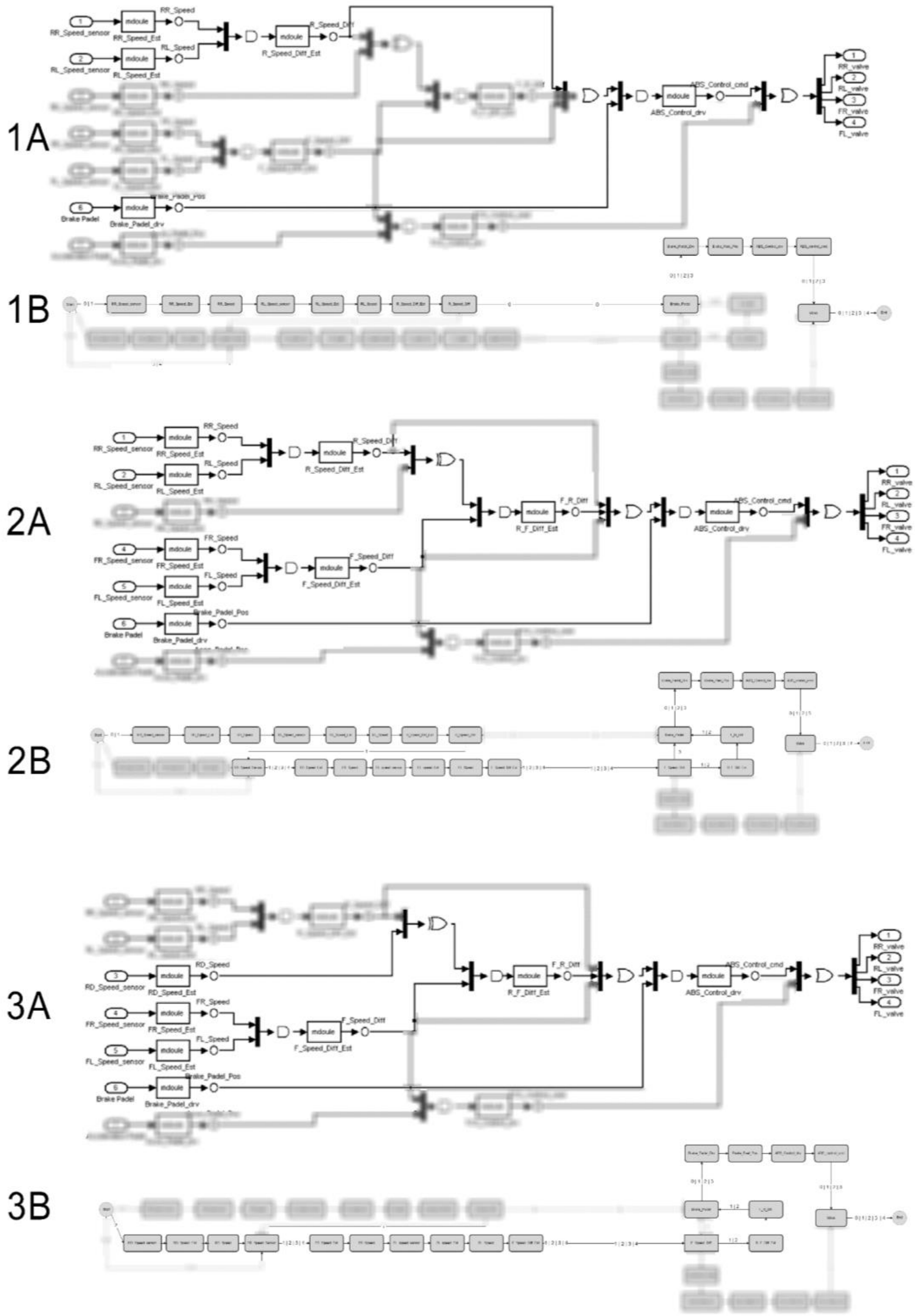

Figure 32. Extracted pipeline of ABS system (1A) first DG solution (1B) sequence for the first path. (2A) Second DG solution (2B) sequence for the second path. (3A) Third DG solution (3B) sequence for the third path. (4A) Fourth DG solution (4B) sequence for the fourth path. (5A) TCS DG solution (5B) sequence for the TCS path.

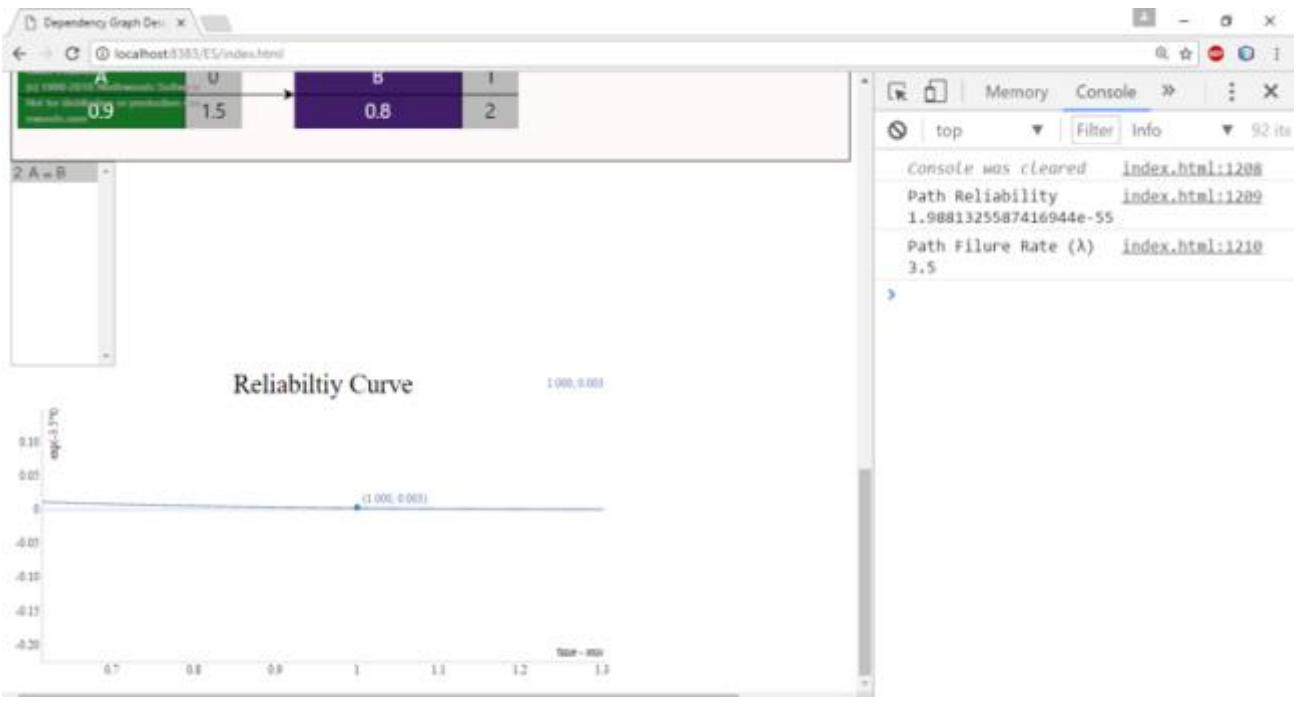

Figure 33. Reliability, failure rate, and a reliability curve appears as results when selecting a pipeline.

## *C. ABS Reliability Analysis*

The reliability of all possible paths for the ABS are shown in Figure 34, where the selected failure rate λ for every node is 1 failure/million hours. In this diagram, Path1 and Path2 overlap, because they have the same number of nodes, Path1 and Path2 together have the optimal solution of reliability of the ABS system. The CAD-tool gives priority to the first pipeline it finds. If the first path fails, then there will be no change on the system reliability level. Path3 comes next in the reliability level and the last will be Path2. The start time is assumed to begin when the ABS gets in duty. The time window for the chart in Figure 34 is 0.2 million hours or 23.41 years.

To analyze the system in more depth, some variations are carried on to study their effect on the reliability of the ABS system. Figure 35 shows the effect of replacing the sensors failure rate to be four times the failure rate of other components. The chart shows that no change happened on the order of paths, Path1 and Path2 still overlap, then comes Path4, and finally Path3, but there is a squeeze in the life time of this ABS system. For example, at time 0.04, the first path's reliability is 0.4, while it was 0.58 in the previous chart illustrated in Figure 34. Figure 36 illustrates the results of the case when the failure rates of the front speed sensors FL_Speed_sensor and FR_Speed_sensor are equal to four times of the failure rate of other components. In this case, Path1 comes first followed by Path4, then comes Path3, then Path2 at the tail.

During the life time of the ABS system in the same case, and for example as illustrated in Figure 37, at time equal to 0.04 million hour (4.68 year), if a failure occurred in Path1 due to a component failure, or a link disconnection, then the ABS system will be degraded to the second reliable Path which is Path2. The drop in ABS reliability will be approximately 0.15 failure/Million hours.

There was another damage to the second path at time 0.05, this causes ABS reliability to decrease again by around 0.1 failure/Million hours from the recent reliability value. When Path1 comes back again, this will allow an abrupt enhance in reliability to become around 0.2 failure/Million hours. Again, at time 0.08 Path1 failed, Path2 is already down, the working path will be Path3. After time 0.1 million hours, Path3 failed, then the Path4 will keep the ABS running, unless a maintenance has carried on for the first paths.

When the fault rate for the rear differential processing element, rear differential data RD_speed data variable, and rear left sensor are five times the fault rate of other components, then the result is shown in Figure 38. Path4 will be the optimal choice, then Path1, Path2 slightly better than Path3.

Figure 39 illustrates the reliability paths of the ABS system when the differential senor failure rate is ten times greater than other components' failure rates. Path1 and Path4 have the same curve at the top, then Path2 follows, and finally comes Path3. From all these graphs, it is clear that changing the arrangements of the failure rate will control the reliability value, and accordingly will affects the process of electing the working path in the ABS system.

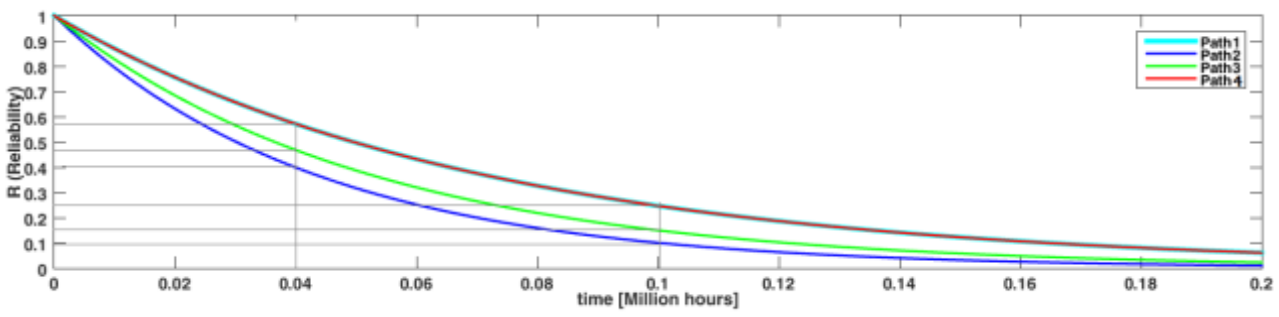

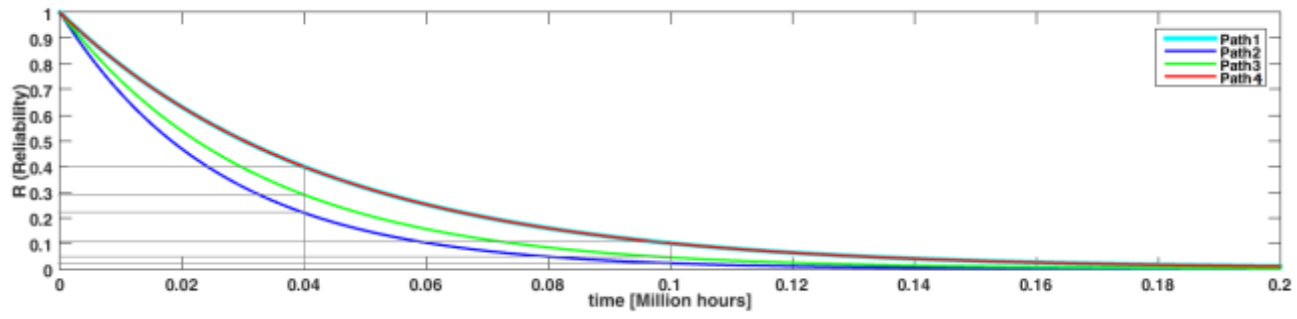

Figure 34. Reliability for ABS Paths when all components have the same failure rate.

Figure 35. Sensors have 4 times the failure rate of other components.

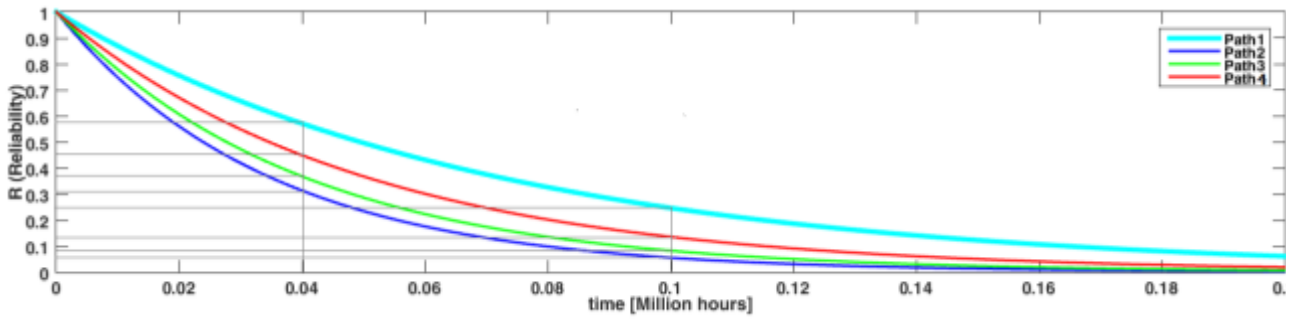

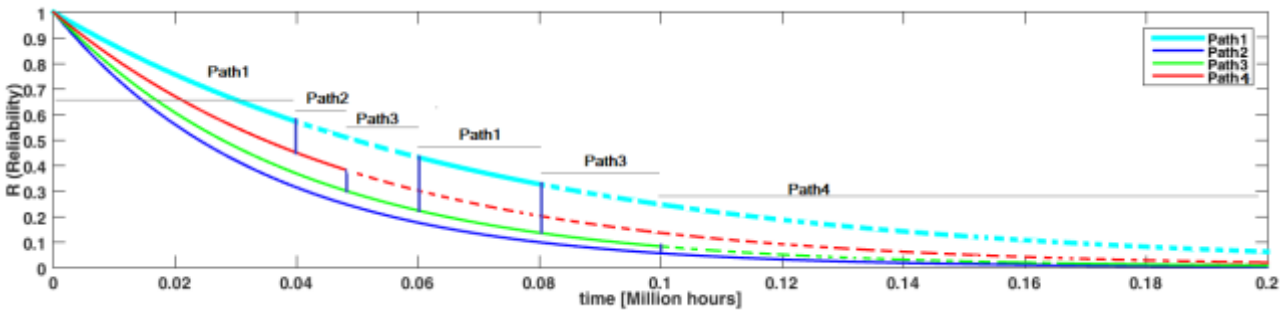

Figure 36. Reliability of the paths when FL_Speed_sensor & FR_Speed_sensor failure rates are 4 times other nodes.

Figure 37. Reliability states of ABS in different failures and maintenance scenario.

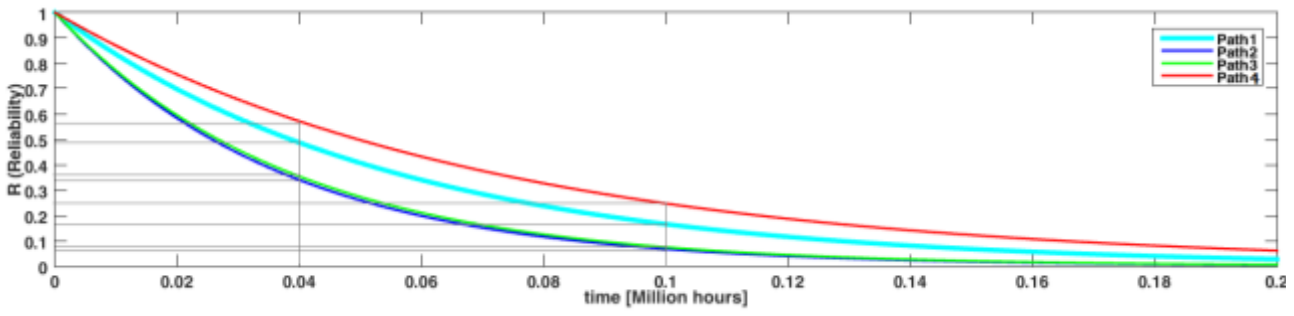

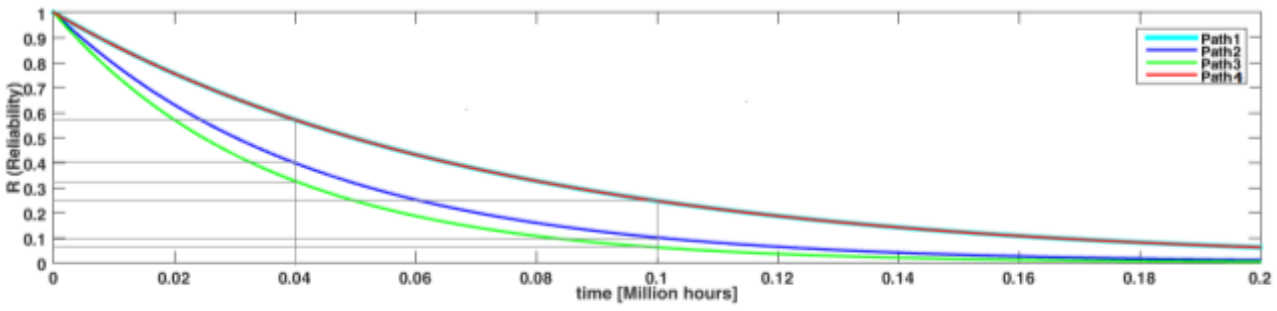

Figure 38. Reliability of ABS when rear differential processing element, rear differential processing element software reliability, and rear left sensor are five times the fault rate of other components.

Figure 39. Reliability of ABS paths when rear differential sensor has 10 times failure rate of others.

### *D. Experimental Implementation on ABS*

Experimental implementation of the CAD-tool in the ABS case according to ARDEA frame work is shown in Figure 40. The CAD-tool in this case is an example of a software-in-the-loop control process. A user can build his own DGs by the CAD-tool or may restore a previously stored program from programs repository. The tool converts the diagram to a data file and vice versa. Then the result of the DG implementation on the CAD-tool will produce a data file. The controller process on the DES system manager is designed to accept importing this data file. According to this data file, the system manager will control the overall PEs in DES through commanding the agents on every PE by the proper command.

As an experimental procedure, the CAD-tool is developed to accept user inputs reliability variables. Input fields include failure rates in terms of failures per one million hours, reliability in a failure per 1 million, name of the node, and IDs in hexadecimal format for the processing elements. Figure 41 shows the full settings and values implementation for the ABS using the CAD-tool.

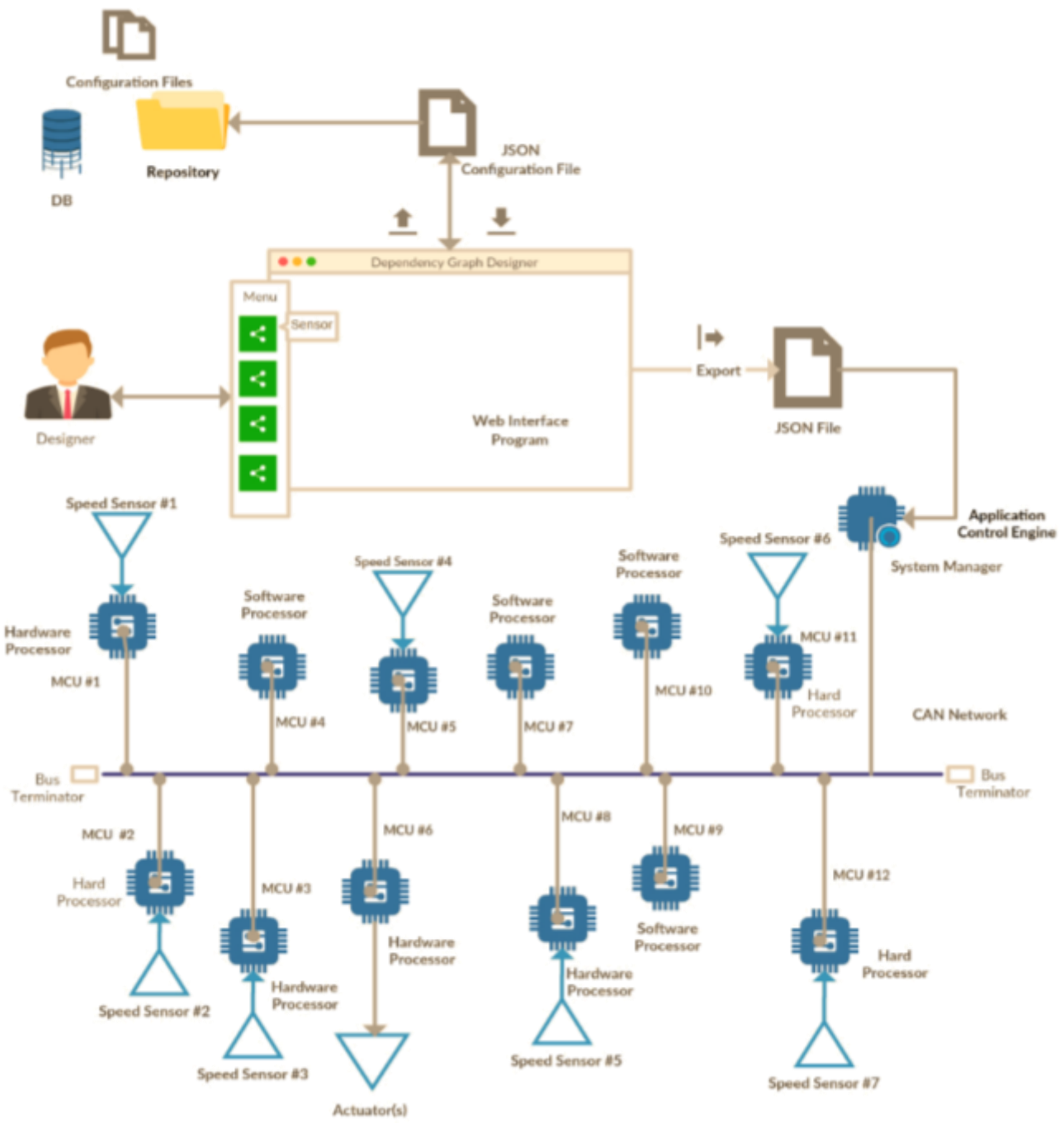

Figure 40. Using the tool to design the FT DES according to ARDEA framework.

Figure 42 illustrates a module node specifications and input fields. Table 2 shows a description of these symbols and numbers. After implementing ABS DES, then the IDs summation for the working PE in every path is shown in Table 3. The JSON file contains this array [{2211, 3003, 2876, 2360}], this array forms the management traffic passed into the system manager via serial communication as shown in Figure 43.

The ABS system manager parses the file and determines what should be chosen as a working path from the status register and the Reliability Options array. Figure 44 illustrates the interaction between the management, control, and forwarding plan of a distributed system composed of three branches with a voter and an actuator shown in Figure 23. The ABS DES is more complex and distributed but it follows the same outlines. Blue lines show the forwarding communication interconnections, gray lines illustrate the control communications, and light red lines in the System Manager circumstance represent the management communication. The hardware required to carry the communication traffic varies, if the communication is for management then it is carried throughout a serial communication channel using RS232 protocol. A serial RS232 driver handles this traffic and provides a channel to the management application task on top of the embedded operating system.

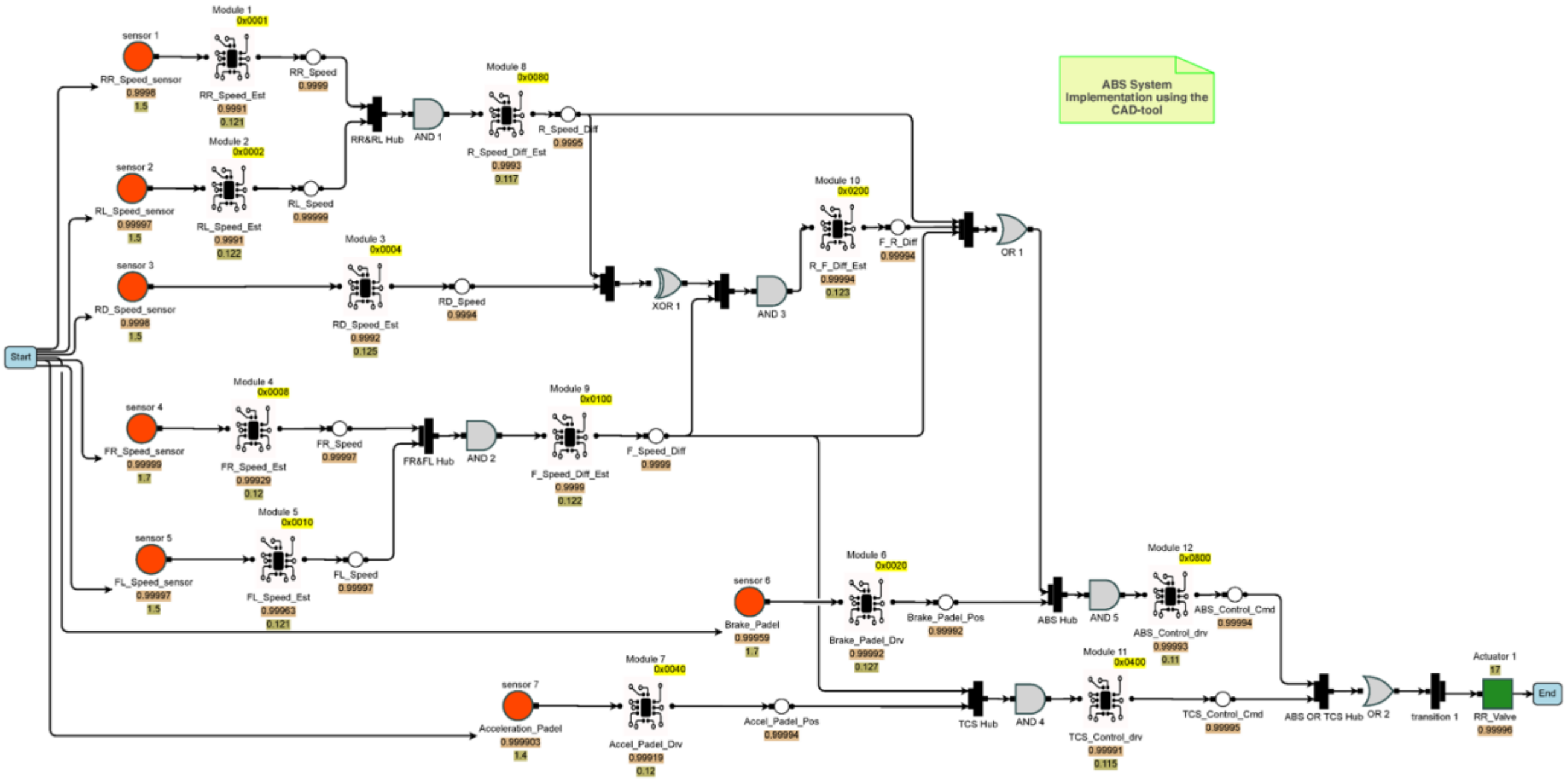

Figure 41. Case study implementation of ABS.

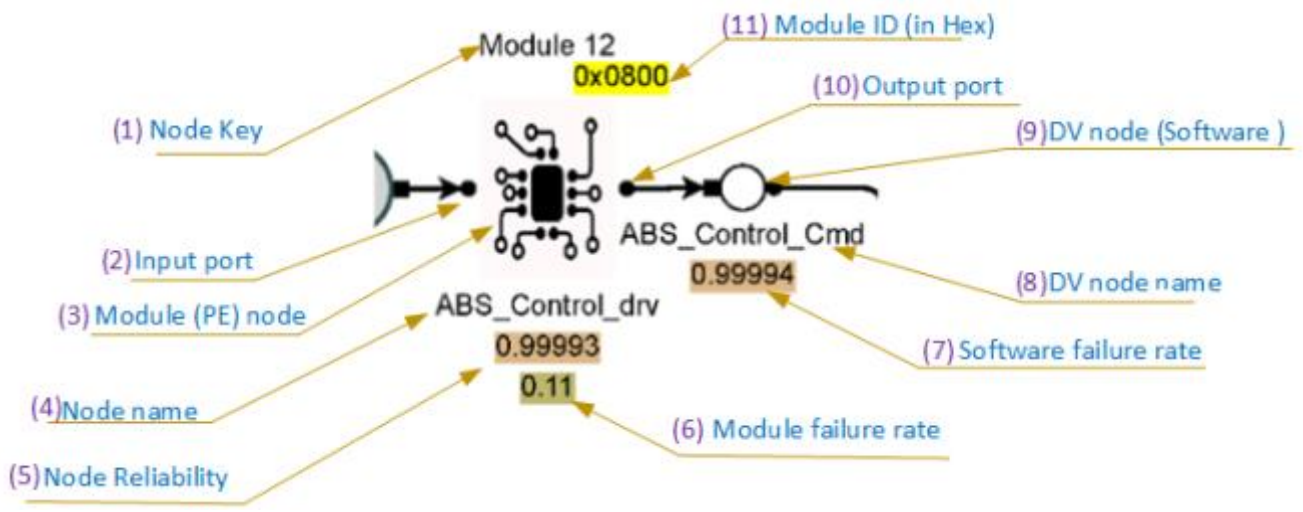

Figure 42. Node specifications and inputs

*TABLE 2*

| Number | Description |
|---|---|
| (1) | Node key is a unique number/text in the DG. |
| (2) | Input port enables connecting links to the node. |
| (3) | Module node is the symbol for the node in DG. |
| (4) | Input field to insert user-defined node name customized for applications. |
| (5) | Node reliability variable input field. |
| (6) | Module node failure rate variable input field. |
| (7) | Input field to insert the failure rate for Module node software (DV). |
| (8) | Input field to insert DV node name. |
| (9) | Software (DV) node symbol. |
| (10) | Output port enables connecting links from the node. |
| (11) | Unique ID number in Hex, number should contain only one binary bit. |

*TABLE 3*

*Summation of IDs for the working PEs*

| Path No. | IDs of the working PEs (Hex) | | | | | | | | | | | | Sum | |
|---|---|---|---|---|---|---|---|---|---|---|---|---|---|---|
| | 1 | 2 | 4 | 8 | 10 | 20 | 40 | 80 | 100 | 200 | 400 | 800 | (hex) | (Dec) |

| | | | | | | | | | | | | |
|---|---|---|---|---|---|---|---|---|---|---|---|---|
| Path1 | ● | ● | | | | ● | ● | | | ● | 8A3 | 2211 |
| Path2 | ● | ● | | ● | ● | ● | ● | ● | ● | ● | BBB | 3003 |
| Path3 | | | ● | ● | ● | ● | | ● | ● | ● | B3C | 2876 |
| Path4 | | | | ● | ● | ● | | ● | | ● | 938 | 2360 |

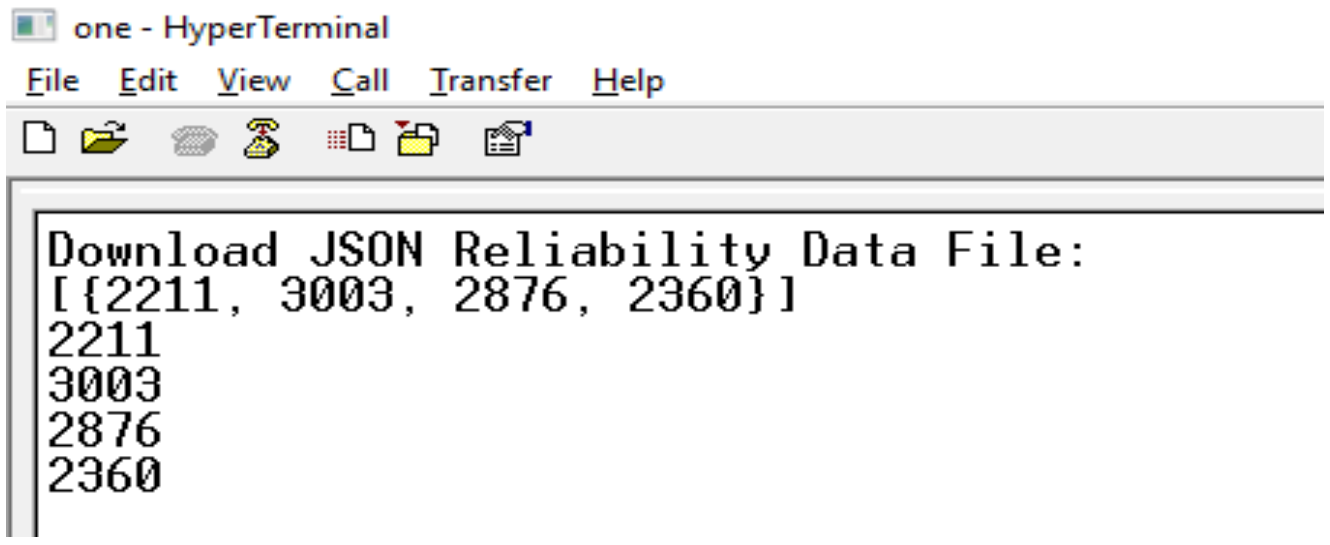

Figure 43. Importing and parsing Reliability Options file by the system manager.

If the traffic is control traffic, then it is driven by the CAN network driver towards the controller application task on the System Manager side, or towards the agent task if it was on the PE side. CAN is also used in case of forwarding traffic passing among application tasks on the different PEs. Applications on one node can communicate with applications running on other nodes. To communicate efficiently across the CAN network, the messages address space should be designed and divided into groups, this will manage the flow of information within the network. Table 4 shows the address space map for the used CAN network. Because the standard version of CAN is used, a total space of message addressing is 2048 different addresses (0-2047). Lower addresses have higher priority. The lowest category of address space is kept for emergency addresses and for the broadcast message from the manager to all nodes in the network. As Table 4 shows, emergency category addresses are from 0 to 127, it is also subdivided into four sub categories, every subcategory includes a range of 32 different addresses. These categories for rationing purposes, some of subcategories are used and others are left for future expansion of the networking system.

TABLE 4

CAN NETWORK MESSAGE ADDRESS SPACE MAP

| **Category** | | **Subcategory beginning** | **Subcategory beginning (Hex)** | **Range** | **Description** |
|---|---|---|---|---|---|
| Emergency | | | | | |
| 0 | | 0 | 0 | 0-31 | |
| | | 32 | 20 | 32-63 | |
| | | 64 | 40 | 64-95 | MANAGER TO ALL |
| | 127 | 96 | 60 | 96-127 | |
| Manager | | | | | |
| 128 | | 128 | 80 | 128-159 | Hardware PE to Manager |
| | | 160 | A0 | 160-191 | Manager PE to Actuators |
| | | 192 | C0 | 192-223 | Manager PE to Software PE |
| | 255 | 224 | E0 | 224-255 | Manager to Hardware PE |
| Software PE | | | | | |
| 256 | | 256 | 100 | 256-287 | SW PE to Manager |
| | | 288 | 120 | 288-319 | SW PE to Actuators |
| | | 320 | 140 | 320-351 | SW PE to Software PE |
| | 383 | 352 | 160 | 352-383 | SW PE to Hardware PE |
| HW PE | | | | | |
| 384 | | 384 | 180 | 384-415 | Hardware PE to Manager |
| | | 416 | 1A0 | 416-447 | Hardware PE to Actuators |

TABLE 4

CAN NETWORK MESSAGE ADDRESS SPACE MAP

| Category | | Subcategory beginning | Subcategory beginning (Hex) | Range | Description |
|---|---|---|---|---|---|
| | | 448 | 1C0 | 448-479 | Hardware PE to Software PE |
| | 511 | 480 | 1E0 | 480-511 | Hardware PE to Hardware PE |
| Actuator | | | | | |
| 512 | | 512 | 200 | 512-543 | Actuator to Manager |
| | | 544 | 220 | 544-575 | |
| | | 576 | 240 | 576-607 | |
| | 639 | 608 | 260 | 608-639 | |
| Block1 | | | | | |
| 640 | | 640 | 280 | 640-671 | |
| | | 672 | 2A0 | 672-703 | |
| | | 704 | 2C0 | 704-735 | |
| | 767 | 736 | 2E0 | 736-767 | |
| Block2 | | | | | |
| 768 | | 768 | 300 | 768-799 | |
| | | 800 | 320 | 800-831 | |
| | | 832 | 340 | 832-863 | |
| | 895 | 864 | 360 | 864-895 | |
| 896 – 2047 | | Unallocated for system expansion | | | |

The selected identifiers of the CAN bus for the ABS system are shown in Table 5. The listed identifiers are used by the network application and CAN driver to identify the relevant messages and pass them to the upper applications. If a message is not relevant, then the application ignores it. To increase performance, a designer can filter irrelevant CAN messages on the hardware driver instead of processing them.

TABLE 0

CAN NETWORK MESSAGE IDENTIFIERS

| CAN Messene Mnemonics | ID (Hex) | CAN Messene Mnemonics | ID (Hex) |
|---|---|---|---|
| ManagerToRR_Speed_Est | E0 | ManagerToRL_Speed_Est | E1 |
| ManagerToR_Speed_Diff_Est | C0 | ManagerToRD_Speed_Est | E2 |
| ManagerToR_F_Diff_Est | C1 | ManagerToFR_Speed_Est | E3 |
| ManagerToFL_Speed_Est | E4 | ManagerToF_Speed_Diff_Est | C2 |
| ManagerToBrake_Padel_Drv | E5 | ManagerToABS_Control_Drv | C3 |
| ManagerToAccel_Padel_Drv | E6 | ManagerToTCS_Control_Drv | C4 |
| ManagerToValves | A0 | RR_Speed_EstToManager | 180 |
| RR_Speed_EstToR_Speed_Diff_Est | 1C0 | RL_Speed_EstToManager | 181 |
| RL_Speed_EstToR_Speed_Diff_Est | 1C1 | R_Speed_Diff_EstToManager | 100 |
| R_Speed_Diff_EstToR_F_Diff_Est | 140 | RD_Speed_EstToManager | 182 |
| RD_Speed_EstToR_F_Diff_Est | 141 | R_F_Diff_EstToManager | 101 |
| R_F_Diff_EstToABS_Control_Drv | 142 | FR_Speed_EstToManager | 183 |
| FR_Speed_EstToF_Speed_Diff_Est | 1C2 | FL_Speed_EstToManager | 184 |
| FL_Speed_EstToF_Speed_Diff_Est | 1C3 | F_Speed_Diff_EstToManager | 102 |
| F_Speed_Diff_EstToABS_Control_Drv | 143 | F_Speed_Diff_EstToTCS_Control_Drv | 144 |
| Brake_Padel_DrvToManager | 185 | Brake_Padel_DrvToABS_Control_Drv | FA |
| ABS_Control_DrvToManager | 103 | ABS_Control_DrvToActuator | 120 |
| Accel_Padel_DrvToManager | 186 | Accel_Padel_DrvToABS_Control_Drv | 1C4 |
| Accel_Padel_DrvToTCS_Control_Drv | 122 | TCS_Control_DrvToManager | 104 |
| TCS_Control_DrvToValve | 121 | ValvesToManager | 200 |
| R_Speed_Diff_EstToABS_Control_Drv | 145 | F_Speed_Diff_EstToR_F_Diff_Est | 146 |
| TCS_Control_DrvToABS_Control_Drv | 147 | | |

### *E. Comparing the proposed CAD-tool with Koren's Fault Tolerance tool*

To show the result of comparing the CAD-tool with another well-known online tool for reliability. A screenshot of Koren's fault tolerance tool [55] is shown in Figure 45. Koren's tool is built to be a simulation tool, in which the user submits the configuration of nodes in the system by addressing numbers and specifying either the node is an input or an output for a block. It is also required to specify the nodes connected with every block. The designer needs to plot his graph on a paper, name the blocks in a strict way only by alphabetical order, then he has to map the alphabets with the actual block to map between the nodes and the naming. An example of using Koren's tool is illustrated in Figure 46. Then in a separate step, the designer needs to give the nodes failure rates. The input method is done in a text area, after the calculation, a basic graph will be shown for the total reliability of the system.

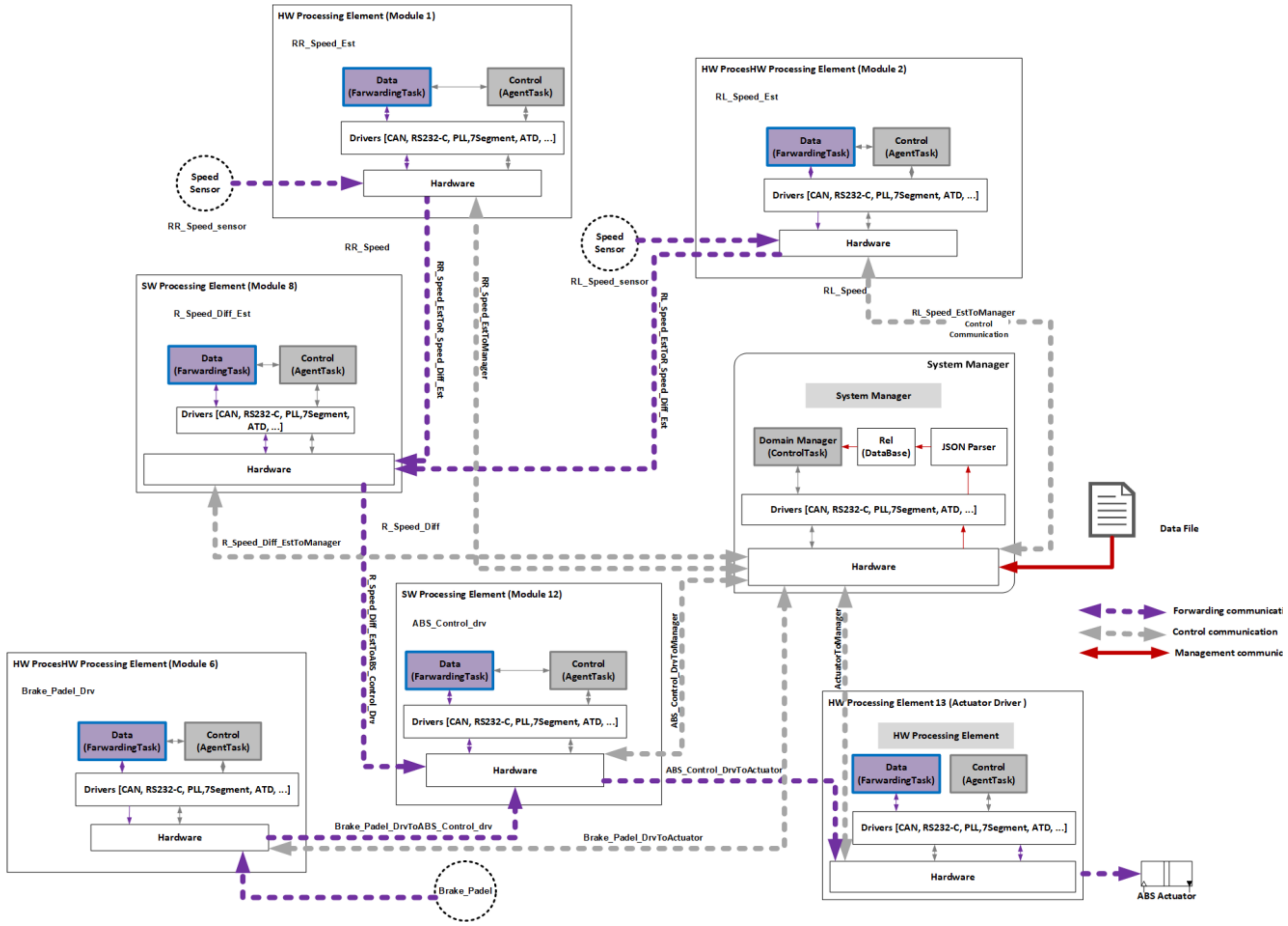

Figure 44. Path1 networking for ABS case study.

In the proposed CAD-tool, the same example is done in Figure 47 but with a simpler way. There is freedom to link any node any time with no restrictions for input conventions, the naming is open, reliability variables and failure rates are written directly on the nodes, interactive interface. The proposed CAD-tool is able to import and export configurations in JSON text files, which enables the scalability of the tool. The reliability curve is not static, interactivity of the plot enables the designer to examine reliability values on progress the curve as shown in Figure 48.

To compare results between our CAD-tool and the known online Koren tool, Table 6 summarizes the differences between both tools.

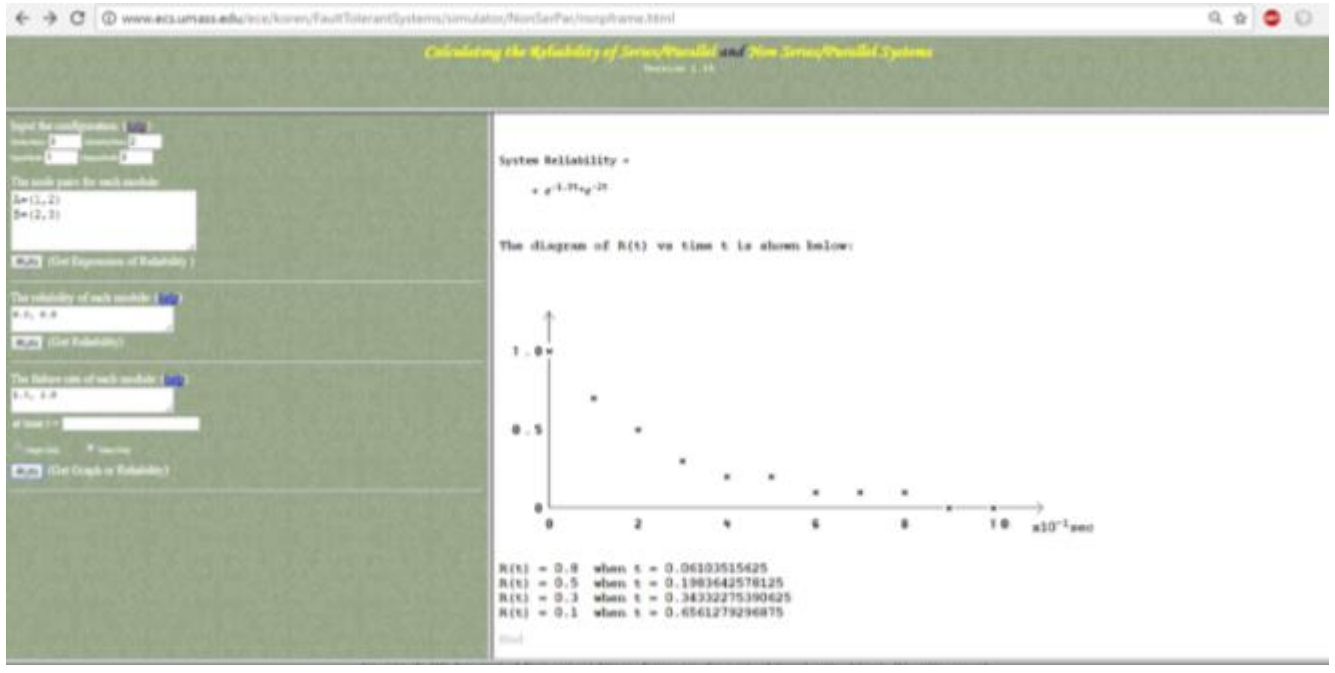

Figure 45. Koren's tool.

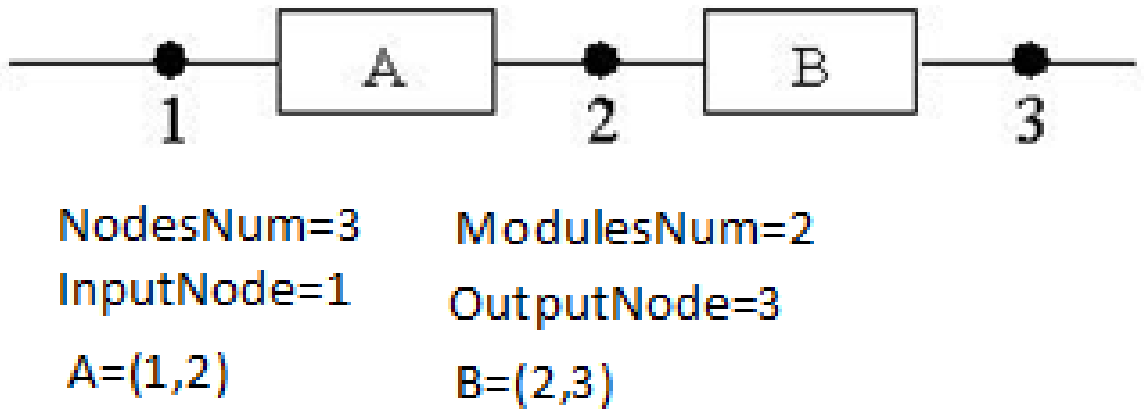

Figure 46. Example of using Koren's tool.

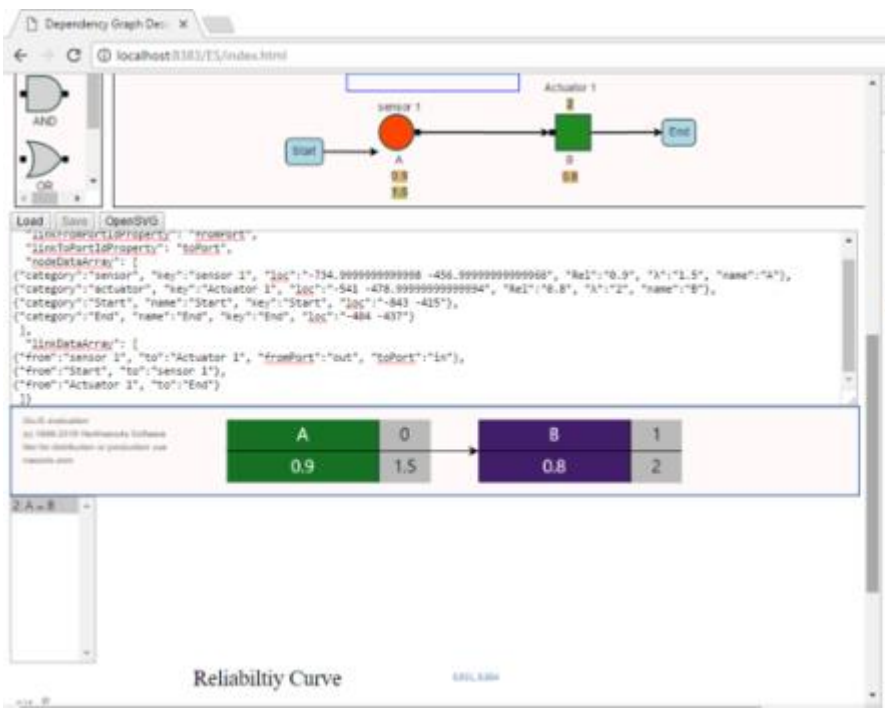

Figure 47. Same Example of Figure 5.22 but by implantation on our CAD-tool.

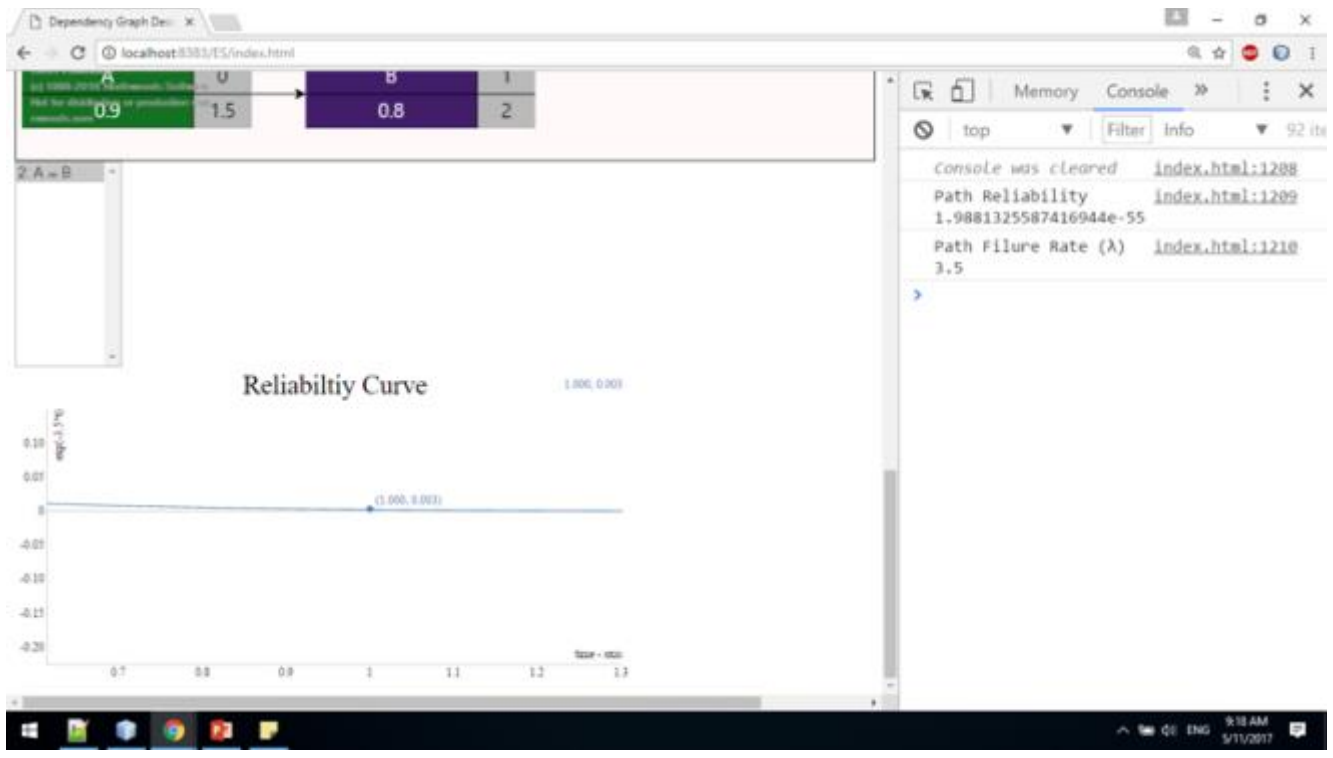

Figure 48. Reliability curve in CAD-tool.

**TABLE 6**

A COMPARISON OF THE CAD-TOOL WITH KOREN'S TOOL

| Koren's Tool | Proposed CAD-tool |
|---|---|
| **Input method is filling fields onside** | Graphical interface inputs |
| **Simulator Only** | Simulator or/and software-in-loop |
| **Node pairs for each module should be specified** | Graphical links specifies interconnections |
| **Node Identifications are very strict (only A, B, ...)** | Very flexible node identifications |
| **No name for a node** | Custom naming |
| **Specifying the reliability of nodes done by filling fields onside.** | Reliability assigned directly on the nodes of DGs |
| **Basic Black and white static graph** | Graph is colored, interactive, zoom-able, and user friendly |
| **Supports limited number of nodes** | Supports unlimited number of nodes |
| **User should check the written syntax before calculations** | No written syntax to check it is graphical |
| **No storing capabilities** | User can backup/restore the configuration file |
| **The tool structure is strict** | Tool can easily be upgraded to bind more functionalities. |
| **Calculates the total reliability of connected nodes** | Extracts the possible sequence of node to form a pipeline |
| **Results are not exported outside the tool** | Results are exported to a JSON file to be used by other systems |

## 6. Conclusion

Embedded computing systems are among the most critical and complex components of today’s technology and industry. They are in most of the safety-critical applications. Such systems need to be dependent, reliable and fault-tolerant. Fault Tolerance design requires more attention from researchers.

In this work, a CAD-based tool to support fault tolerance design in distributed embedded systems was developed. The proposed CAD-tool is based on the theoretical framework called ADREA by Rawashdeh and Lumpp [2].

The tool improves the efficiency of Fault Tolerance design by solving the fault tolerance analysis and complexity. The CAD-based tool uses fault tolerance techniques to determine the best path with the least negative impact in case of a failure. The proposed tool searches, prepares, and sorts the list of possible solutions based on their reliability. Moreover, it is user friendly and easy to use.